\documentclass[a4,11pt]{article}

\usepackage{amsmath,amssymb,amsfonts,color,graphics,graphicx,amscd,amsfonts}
\newcommand{\tr}{\mathop{\mathrm{tr}}}

\def\Tr{\mbox{ Tr }}

\allowdisplaybreaks[4]

\date{}
\begin{document}

\vspace{0.1cm}

\begin{center}
  {\LARGE
  Large-$N_c$ equivalence and 
the sign problem at finite baryon density
}
\end{center}
\vspace{0.1cm}
\vspace{0.1cm}

\begin{center}

          Masanori H{\sc anada} 
          
\vspace{0.5cm}

{\it Department of Physics, University of Washington,

 Seattle, WA 98195-1560, USA}\\
 
 and

 {\it KEK Theory Center, High Energy Accelerator Research Organization (KEK), \\
		Tsukuba 305-0801, Japan}\\
		\vspace{1cm}

\end{center}

\vspace{1.5cm}

\begin{center}
  {\bf Abstract}
\end{center}

QCD with a finite baryon chemical potential, despite its importance, is not well understood because the standard 
lattice QCD simulation is not applicable due to the sign problem.  
Although QCD-like theories which do not suffer from the sign problem have been studied intensively, 
relation to QCD with a finite baryon chemical potential was not clear. 
This paper introduces large-$N_c$ equivalences between QCD and various QCD-like theories. 
These equivalences lead us to a unified viewpoint for QCD with baryon and isospin chemical potentials, $SO(2N_c)$ and $Sp(2N_c)$ gauge theories,   
QCD with adjoint matters and two-color QCD. 
In particular  QCD with the baryon chemical potential is large-$N_c$ equivalent to  its phase quenched version 
in a certain parameter region, 
which is relevant for heavy ion collision experiments. 
All previous simulation results which study the effect of the phase confirm the phase quench approximation is quantitatively good already at $N_c=3$; 
it is so good that often two theories give the same value within error.   
Therefore the phase quenched simulation is the best strategy for the QCD critical point search. 
At small volume one can study a tiny $1/N_c$ effect by the phase reweighting; 
the large-$N_c$ equivalence guarantees that the phase reweighing method works without suffering from the overlapping problem. 


\newpage

\section{Introduction}
\hspace{0.51cm}
Consider QCD at a finite baryon chemical potential (QCD$_B$), 
\begin{eqnarray}
{\cal L} = \frac{1}{4 g^{2} } \tr ({F}_{\mu \nu})^2
+ 
\sum_{f=1}^{N_f} 
\bar{\psi}_{f }\left( \gamma^{\mu} {D}_{\mu} + m_f + \mu \gamma^4 \right)\psi_{f }, 
\end{eqnarray}
where the gauge group is $SU(3)$, $N_f$ is the number of flavors, $\psi_f$ are quarks of mass $m_f$ in the fundamental representation, 
and $\mu$ is the quark chemical potential which is related to the baryon chemical potential $\mu_B$ 
as $\mu_B = 3\mu$.  
Properties of this theory have long been a subject of intense interest. (For a review, see \cite{Fukushima:2010bq}.)  
Apart from its intrinsic theoretical appeal, this subject is important in astrophysics, especially in the study of neutron stars.   
The behavior of QCD$_B$ at asymptotically large $\mu_{B}$ is well understood theoretically due to the asymptotic freedom, 
and QCD$_B$ becomes a color superconductor as $\mu_{B} \rightarrow \infty$~\cite{Alford:2007xm}.  

At more  phenomenologically realistic densities, QCD is strongly coupled, and thus not amenable to controlled analytic treatment.  
Although lattice Monte Carlo is very useful at $\mu_{B}=0$,  
however, it runs into trouble at $\mu_{B}\neq 0$ due to the {\it fermion sign problem} -- 
the fermion determinant $\prod_{f=1}^{N_f}\det \left( \gamma^{\mu} {D}_{\mu} + m_f + \mu \gamma^4 \right)$ 
becomes complex, rendering importance sampling exponentially difficult.

In order to circumvent this difficulty people have studied gauge theories which do not suffer from the sign problem at finite density. 
Consider QCD and QCD-like theories\footnote{
In this paper we call $SU(N_c)$ Yang-Mills with $N_f$ fundamental fermions `QCD'. 
$SU(N_c)$ Yang-Mills with fermions in other representations and $SO(2N_c)$/$Sp(2N_c)$ theories 
are referred to `QCD-like theories'.  
} of the form 
\begin{eqnarray}
{\cal L}_{\rm G} = \frac{1}{4 g_{\rm G}^{2} } \tr ({F}^{\rm G}_{\mu \nu})^2
+ 
\sum_{f=1}^{N_f} 
\bar{\psi}^{\rm G}_{f }\left( \gamma^{\mu} {D}^G_{\mu} + m_f + \mu_f \gamma^4 \right)\psi^{\rm G}_{f }, 
\label{QCDlike_action}
\end{eqnarray}
where $G$ is the gauge group e.g. $SU(N_c)$, 
$\mu_f$ is a generic quark chemical potential, and fermions $\psi^{\rm G}$ are not necessarily in the fundamental representation. 
The main examples are QCD with an isospin chemical potential $\mu_I$ (i.e. $N_f$ is even, $\mu_1=-\mu_2=\mu_3=-\mu_4=\cdots =\mu_I/2$; we call this theory as QCD$_I$) 
and degenerate mass \cite{Son:2000xc}, 
two-color QCD of even number of flavors and degenerate mass \cite{Kogut:1999iv}, QCD with adjoint fermions \cite{Kogut:1999iv}, 
and $SO(2N_c)$ and $Sp(2N_c)$ Yang-Mills theories \cite{Cherman:2010jj,Hanada:2011ju}.  
However, while interesting, these theories have many qualitative differences from $N_{c}=3$ QCD, 
such as {\it e.g.} explicitly broken flavor symmetry in the first case. Therefore it is important to understand  
{\it what we can learn from these theories}, or in other words, {\it in what sense they are similar to real QCD with the baryon chemical potential}.  

In \cite{Cherman:2010jj,Hanada:2011ju}, 
an answer to this question has been given. 
The statements are 
\begin{itemize}
\item

$SO(2N_c)$ and $Sp(2N_c)$ theories with fundamental fermions and $\mu_B$ (SO$_B$ and Sp$_B$) and QCD$_I$ are large-$N_c$ equivalent 
both in the 't Hooft limit ($N_f$ fixed) and the Veneziano limit ($N_f/N_c$ fixed), everywhere in the $T$-$\mu$ plane. 

\item
SO$_B$, Sp$_B$, QCD$_I$ 
and QCD$_B$ are large-$N_c$ equivalent 
in the 't Hooft limit, outside the BEC/BCS crossover region of the former three theories. 

\item
More generally, $SO(2N_c)$, $Sp(2N_c)$ and $SU(N_c)$ theories with fermion mass $m_1,\cdots,m_{N_f}$ and 
chemical potential $\mu_1,\cdots,\mu_{N_f}$ are equivalent. The signs of the quark chemical potentials can be flipped 
without spoiling the equivalence. (Fig.~\ref{fig:web_general}) 

\item
$SO(2N_c)$ YM with the $N_f$  complex adjoint fermions and $\mu_B$, 
$SU(N_c)$ YM with the $N_f$  complex adjoint fermions and $\mu_B$, 
and $SU(N_c)$ YM with the $2N_f$  complex anti-symmetric fermions and $\mu_I$ 
are large-$N_c$ equivalent  everywhere in the $T$-$\mu$ plane. 

\end{itemize}
Furthermore, 
\begin{itemize}
\item
At finite-$N_c$, QCD$_B$ and QCD$_I$ agree up to $(N_f/N_c)^2$ corrections. Previous simulation results 
show the agreement is very good already at $N_c=3$. 

\item
One can take into account $1/N_c$ corrections by the phase reweighting method. 
The large-$N_c$ equivalence guarantees the absence of a severe overlapping problem. 

\end{itemize}

\begin{figure}[htbp]
\begin{center}
\scalebox{0.7}{
\includegraphics[width=10cm]{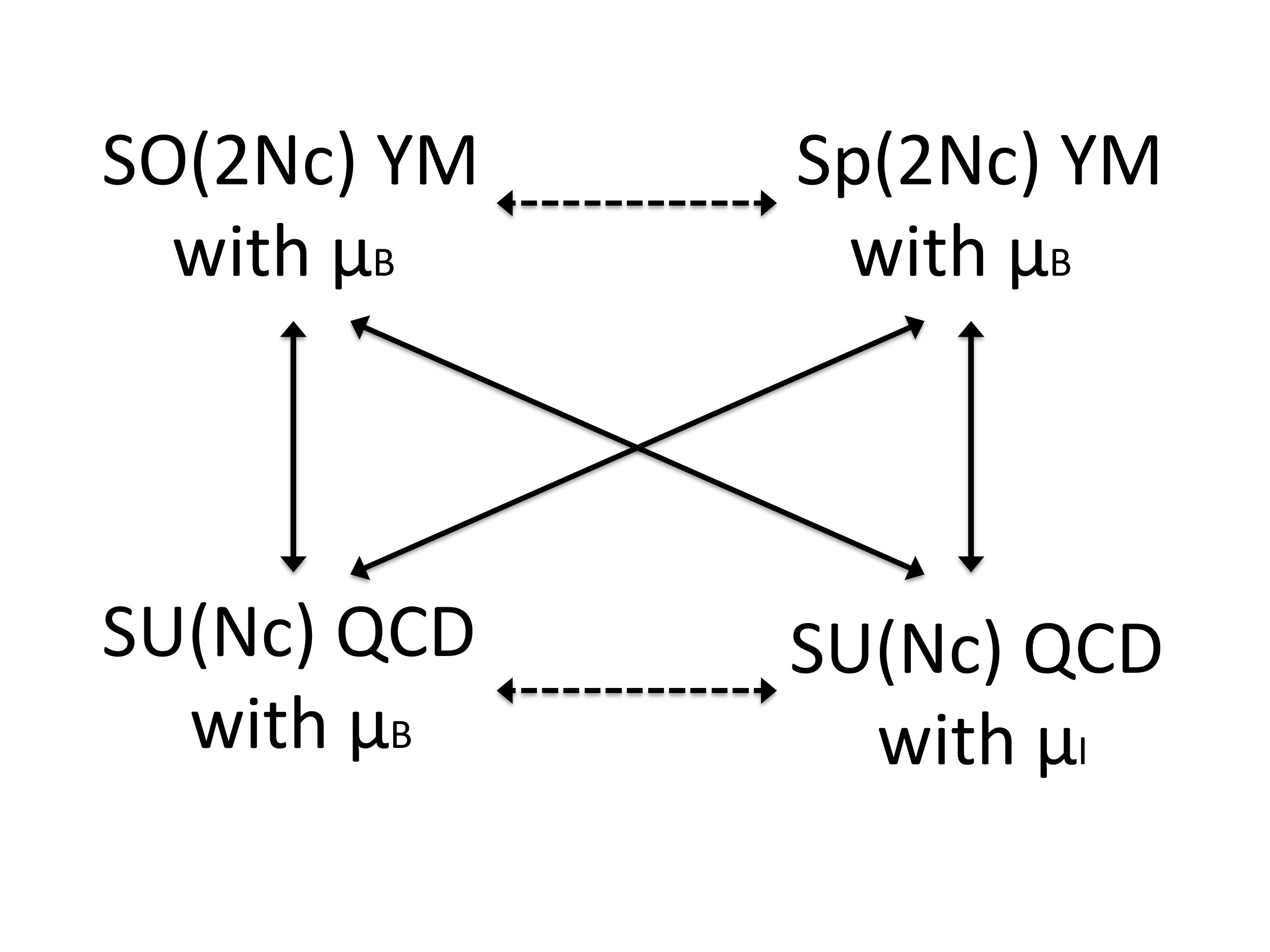}}
\end{center}
\vspace{-0.5cm}
\caption{A web of equivalences. Arrows with solid lines represents equivalences through orbifold projections. 
Arrows with dashed lines are the `parent-parent' and `daughter-daughter' equivalences which arise as combinations 
of the orbifold equivalences.  
}
\label{fig:web}
\end{figure}

\begin{figure}[htbp]
\begin{center}
\scalebox{0.7}{
\includegraphics[width=10cm]{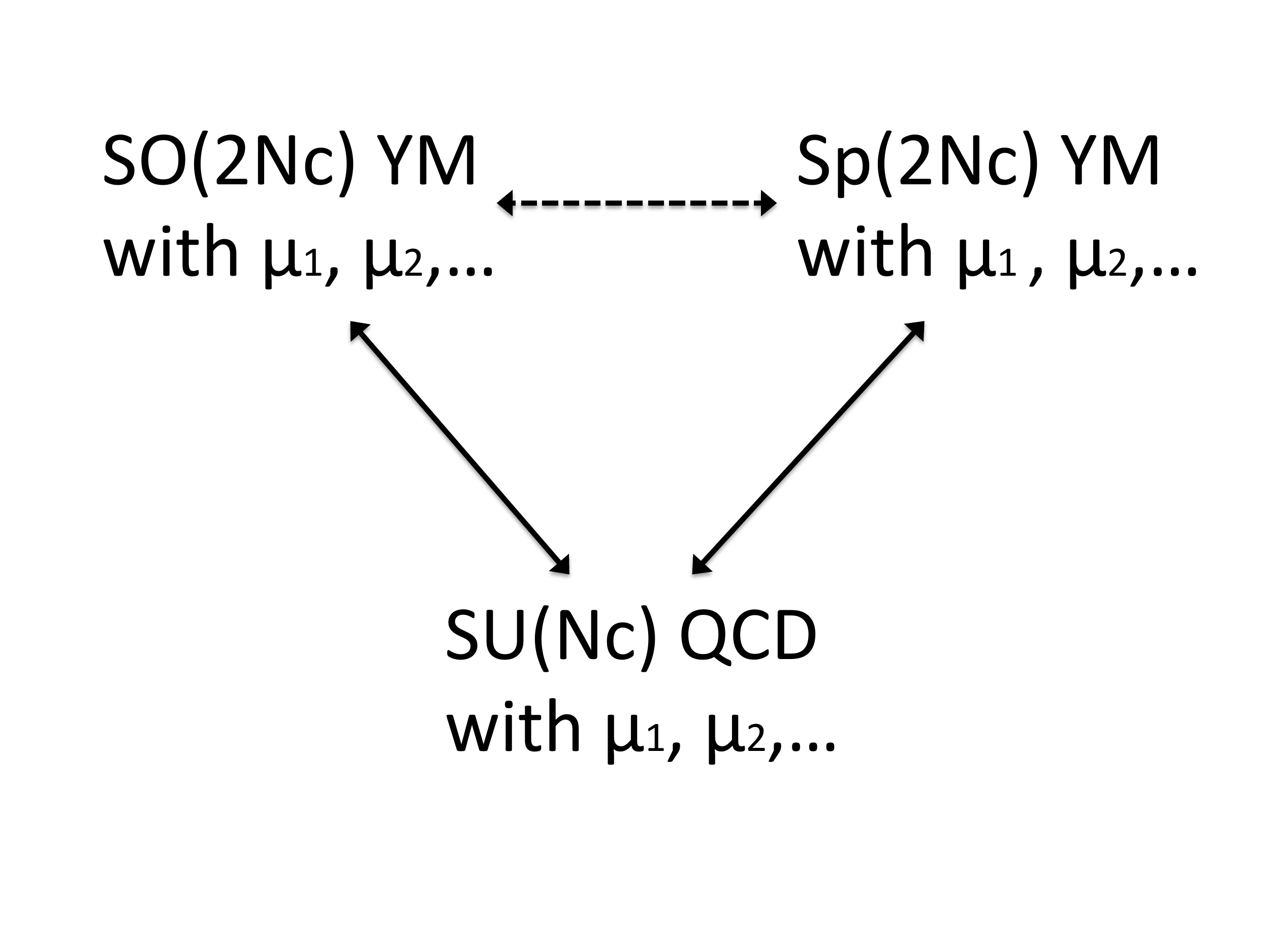}}
\end{center}
\vspace{-0.5cm}
\caption{More general version of the equivalences. 
Values of the quark chemical potentials can be different. 
}
\label{fig:web_general}
\end{figure}

\begin{figure}[htbp]
\begin{center}
\scalebox{0.7}{
\includegraphics[width=10cm]{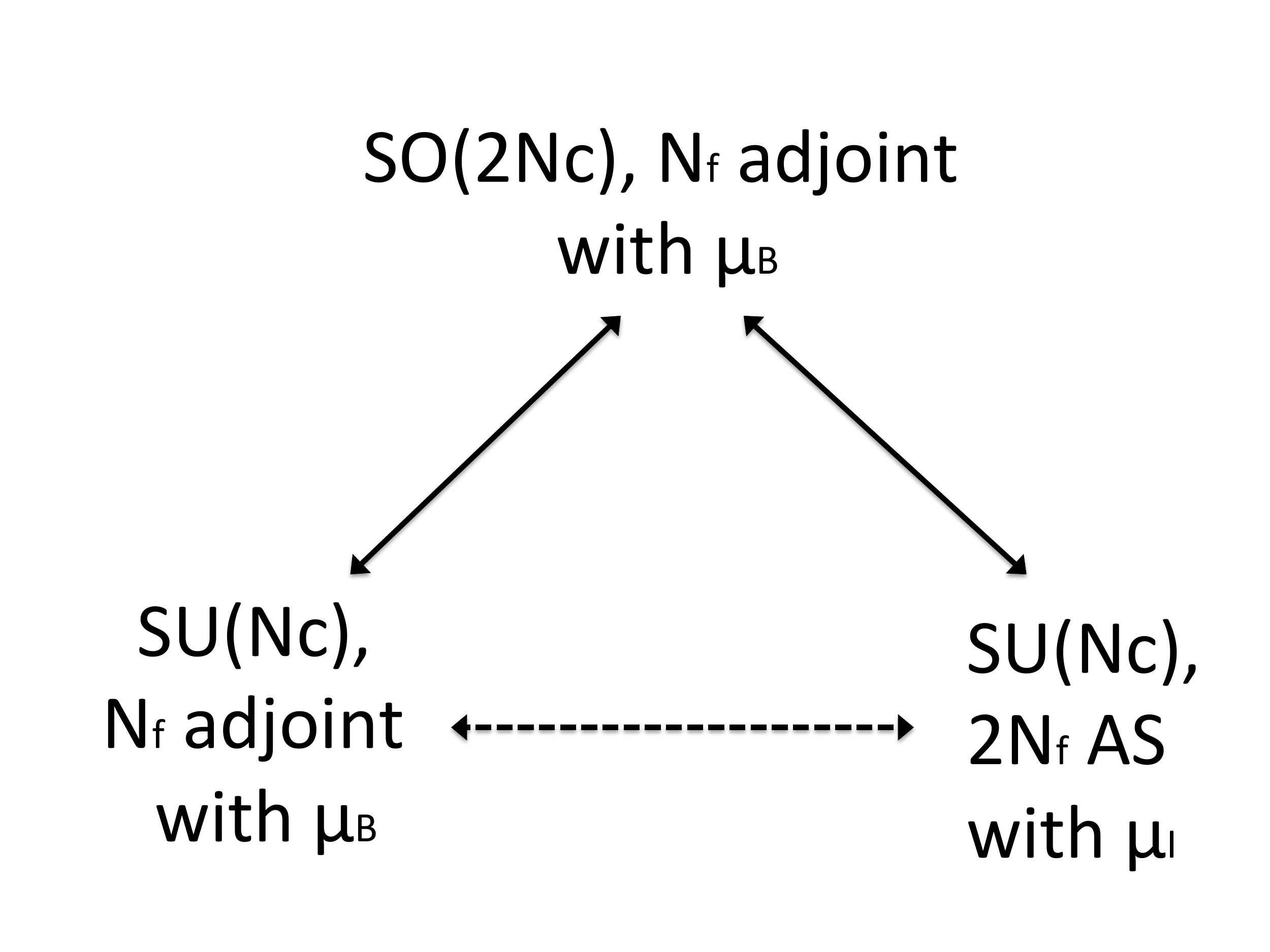}}
\end{center}
\vspace{-0.5cm}
\caption{ Equivalences in the Corrigan-Ramond limit. $SU(N_c)$ YM with anti-symmetric fermions 
can be regarded as a special kind of large-$N_c$ limit of three-color QCD (the Corrigan-Ramond limit), because 
anti-symmetric and fundamental representations are equivalent at $N_c=3$. 
Unfortunately, $SU(N_c)$ YM with anti-symmetric fermions and $\mu_B$ 
cannot be incorporated in these equivalences. 
}
\label{fig:web2}
\end{figure}

These statements have been derived by using string-inspired large-$N_c$ techniques  
\cite{Kachru:1998ys,Bershadsky:1998mb,Bershadsky:1998cb,Kovtun:2003hr}.
As shown in \cite{Cherman:2010jj,Hanada:2011ju}, there are orbifold projections relating 
SO$_B$, Sp$_B$, 
QCD$_B$ and QCD$_I$ (Fig.~\ref{fig:web}). 
At large-$N_c$, the orbifold equivalence 
guarantees these theories are equivalent in the sense a class of correlation functions (e.g. the expectation value of the chiral condensate and 
$\pi^0$ correlation functions) 
and the phase diagrams determined by such quantities coincide, 
as long as the projection symmetry is not broken spontaneously \cite{Kovtun:2003hr}. 
A similar argument shows QCD with adjoint fermions and $\mu_B$ 
is equivalent to QCD with fermions of two-index antisymmetric representation, which is a so-called the Corrigan-Ramond large-$N_c$ limit, 
with $\mu_I$ (Fig.~\ref{fig:web2}).  
In order for these equivalence to hold, orbifolding symmetries must not be broken spontaneously. 
This requirement is always satisfied for the equivalences between 
SO$_B$, Sp$_B$ and QCD$_I$. 
For the equivalences between these three theories and QCD$_B$,  
`outside the BEC/BCS crossover region' is required for the symmetry realization. 
This region (`outside the BEC/BCS crossover region') is relevant for the search for the {\it QCD critical point}, 
which attract intense interest over the decade.  
Our answer to the problem is strikingly simple -- by using sign-free theories 
one can answer to the question. 
In the case of the two-flavor theory, QCD$_I$ is nothing but the phase-quenched version of QCD$_B$. 
Therefore, the sign problem is merely an illusion, up to the $1/N_c$ correction. 
In fact this fact has been known empirically, as nicely summarized for example in \cite{Kogut:2007mz}.  
Actually {\it all previous simulation results which study the effect of the phase confirm the phase quench approximation is quantitatively very good already at $N_c=3$}; 
{\it the agreement is so good that often two theories are indistinguishable}.\footnote{
This is the answer to the criticism from referees, `The author must prove the effectiveness of the method.' 
} 
This is not surprising at all: for flavor physics 
the $1/N_c$-expansion is quantitatively good at $N_c=3$, 
often much better than a naive expectation, and that is the reason why it has been studied since 't Hooft.   
In \S~\ref{sec:SU(3)} we review some previous simulation results which prove the validity of the phase quenching at $N_c=3$. 
Note that the $1/N_c$ correction can be studied by the phase reweighting; 
the orbifold equivalence guarantees that the overlapping problem can be avoided by using the phase quenched ensemble, 
which provides us with the first solution to the overlapping problem with a full theoretical justification.

This paper is organized as follows. 
In \S~\ref{sec:sign} we show the absence of the sign in SO$_B$ and Sp$_B$. 
In \S~\ref{sec:equivalence}, we review the orbifold equivalence and provide a proof to all order in perturbation\footnote{
We do not show the proof the equivalence between theories with adjoint and antisymmetric fermions.  
The proof for the theories with fundamental fermion can easily be modified for this case. 
The projection is the same as the one used in \cite{Unsal:2006pj}.   
The equivalence between two $SU(N_c)$ theories (when the chemical potential is zero) has been found in \cite{Armoni:2003gp}. 
}.  
As discussed in \cite{Kovtun:2003hr}, projection symmetries must be unbroken in order for the equivalences to hold. 
In order to see the fate of the symmetries, we discuss the phase diagrams of SO$_B$, Sp$_B$, QCD$_I$ and QCD$_B$ in \S~\ref{sec:phase}.  
In \S~\ref{sec:SU(3)} we review previous simulation results which studied the effect of the phase and show that the phase quench gives 
qualitatively good answer. 
In \S~\ref{sec:analytic} we generalize the equivalences to analytically solvable toy models 
and confirm the equivalence explicitly. Results in this section strongly suggests the validity of the equivalences 
at nonperturbative level. \S~\ref{sec:conclusion} is devoted for the conclusion and outlook. 

This paper has been prepared for Seitaro Nakamura prize competition, and is based on papers \cite{Cherman:2010jj,Hanada:2011ju,HMY12}  
and a few new results. In 2011 the first version of this paper was rejected, with referees' comment ``The author must prove the effectiveness of the method." \footnote{
More precisely: ``There is a very interesting paper among the ones which were not chosen this time. We encourage him to challenge again, after proving the effectiveness of the method. For such a difficult problem studied over decades, for which many solutions had been proposed, it is very important to prove the effectiveness."
}
Therefore we have added a new section,  \S~\ref{sec:SU(3)}, in which we pointed out that  
all previous simulations which study the effect of the phase confirm the phase quench approximation is quantitatively very good already at $N_c=3$; 
the agreement is so good that often two theories completely agree within numerical error.  

\subsection*{Note added after the second rejection}
\hspace{0.51cm}
This paper had been rejected again in 2012, this time with a comment ``The applicant must realize that the paper is not an application document written for referees. Eventually it will be published." 

\section{Absence of the sign problem in $SO(2N_c)$ and $Sp(2N_c)$ theories}\label{sec:sign}
\hspace{0.51cm}
In this section we prove the absence of the sign problem in SO$_B$ and Sp$_B$. 
Let us start with $SU(N_c)$ theory. 
In four-dimensional Euclidean space, the gamma matrices can be taken Hermitian, 
$
\gamma_\mu^\dagger=\gamma_\mu
$. 
The covariant derivative $D^{\rm SU}_\mu=\partial_\mu+iA_\mu^{\rm SU}$, where 
$A_\mu^{\rm SU}$ is an $N_c\times N_c$ Hermitian matrix, is skew-hermitian, 
$
\left(D^{\rm SU}_\mu\right)^\dagger=-D^{\rm SU}_\mu
$. 
Therefore, $\gamma_\mu D_\mu$ is also skew-Hermitian, 
\begin{eqnarray}
\left(\gamma^\mu D^{\rm SU}_\mu\right)^\dagger
=
-\gamma^\mu D^{\rm SU}_\mu. 
\end{eqnarray} 
Hence the eigenvalues of $\gamma^\mu D^{\rm SU}_\mu$ are pure imaginary. 
Furthermore the eigenvalues appear in pairs $\pm i\lambda$, where $\lambda$ is real, 
because of the chiral symmetry $\gamma_5\left(\gamma^\mu D^{\rm SU}_\mu\right)\gamma_5=-\gamma^\mu D^{\rm SU}_\mu$. 
When the mass $m_f$ is turned on, the eigenvalues are shifted to $\pm i\lambda + m_f$.  
Hence the eigenvalues appear with their complex conjugate (as long as $m_f$ is real) and hence the determinant, 
which is the product of the eigenvalues, is real positive.  

Once the chemical potential $\mu$ is turned on, the skew-Hermiticity is lost,  
\begin{eqnarray}
\left(\gamma^\mu D^{\rm SU}_\mu+\mu\gamma_4\right)^\dagger
=
-\gamma^\mu D^{\rm SU}_\mu+\mu\gamma_4,  
\end{eqnarray} 
and hence the determinant is complex in general. 
Note however that 
\begin{eqnarray}
\left[\det \left( \gamma^{\mu} {D}_{\mu} + m+ \mu\gamma^4 \right)\right]^\ast
=
\det \left( \gamma^{\mu} {D}_{\mu} + m - \mu\gamma^4 \right). 
\end{eqnarray}
For this reason QCD$_I$ with degenerate mass is sign-free; 
the determinant is 
\begin{eqnarray}
\det \left( \gamma^{\mu} {D}_{\mu} + m + \mu \gamma^4 \right)\times \det \left( \gamma^{\mu} {D}_{\mu} + m - \mu \gamma^4 \right)
=
\left|\det \left( \gamma^{\mu} {D}_{\mu} + m + \mu \gamma^4 \right)\right|^2 \ge 0. 
\end{eqnarray}

In $SO(2N_c)$ theory, the sign is absent thanks to an additional anti-unitary symmetry. 
The crucial point is $D_\mu^{\rm SO}$ is real in the coordinate basis; both $\partial_\mu$ and $iA_\mu^{\rm SO}$ are 
real antisymmetric. Therefore 
\begin{eqnarray}
(C\gamma_5)
\left( \gamma^{\mu} {D}_{\mu} + m+ \mu\gamma^4 \right)
(C\gamma_5)^{-1}
=
\left( \gamma^{\mu} {D}_{\mu} + m+ \mu\gamma^4 \right)^\ast, 
\end{eqnarray}
where $C$ is the charge conjugation matrix satisfying $C\gamma_\mu C^{-1}=-\gamma_\mu^T=-\gamma_\mu^\ast$. 
It guarantees the pair structure of the eigenvalues $(\lambda, \lambda^\ast)$; 
if $v$ is an eigenvector with an eigenvalue $\lambda$, $(C\gamma_5)^{-1}v^\ast$ is another eigenvector with an eigenvalue 
$\lambda^\ast$, and furthermore, they are linearly independent even when $\lambda$ is real \cite{Cherman:2010jj}. 

In $Sp(2N_c)$ theory, a similar relation holds, 
\begin{eqnarray}
(J_cC\gamma_5)
\left( \gamma^{\mu} {D}_{\mu} + m+ \mu\gamma^4 \right)
(J_cC\gamma_5)^{-1}
=
\left( \gamma^{\mu} {D}_{\mu} + m+ \mu\gamma^4 \right)^\ast, 
\end{eqnarray}
where $J_c = i\sigma_{2} \otimes 1_{N_{c}}$ (see \S~\ref{sec:Sp}). 
Unlikely to the case of $SO(2N_c)$, this relation does not give the pair structure when $\lambda$ is real, 
and hence only the reality, not the positivity, is guaranteed.  
However as long as the mass and chemical potential are degenerate ($m_1=m_2$, $m_3=m_4$, $\cdots$, $\mu_1=\mu_2$, $\mu_3=\mu_4$, $\cdots$) 
the determinant is real and positive.

\section{Orbifold equivalence}\label{sec:equivalence}
%
\subsection{Pure Yang-Mills theory}\label{proof:pureYM}
\subsubsection{$SO(2N_c)$ to $SU(N_c)$}
\hspace{0.51cm}
The notion of the orbifold equivalence came from the string theory \cite{Kachru:1998ys,Bershadsky:1998mb}. 
Soon it has been proven by using only field theory techniques 
\cite{Bershadsky:1998cb,Kovtun:2003hr}, without referring to the string theory.  
As a simple example, let us consider the equivalence between $SO(2N_c)$ and $SU(N_c)$ pure Yang-Mills theories. 
To perform an orbifold projection, one identifies a discrete subgroup of the symmetry group of the `parent' theory, 
which is the $SO(2N_{c})$ theory in this case, and require the fields to be invariant under the discrete symmetry.  
This gives a `daughter' theory, which is $SU(N_{c})$ YM.  
The orbifold projection uses a $\mathbb{Z}_{2}$ subgroup of the $SO(2N_{c})$ gauge symmetry.   

Let us take $J_c\in SO(2N_{c})$ to be $J_c = i\sigma_{2} \otimes 1_{N_{c}}$, which generates a $\mathbb{Z}_{4}$ subgroup of $SO(2N_{c})$. 
Here $1_{N}$ is an $N \times N$ identity matrix.    
We require the gauge field $A_\mu$ to be invariant under 
\begin{eqnarray}
A_{\mu} \rightarrow J_c A_{\mu} J_c^{-1},  \label{projection_gauge_field}
\end{eqnarray}
which generates a $\mathbb{Z}_{2}$ subgroup of $SO(2N_{c})$. 
$A_{\mu}$ can be written in $N_{c}\times N_{c}$ blocks as
\begin{align}
A_\mu
=
i\left(
\begin{array}{cc}
A_\mu^A+B_\mu^A & C_\mu^A-D_\mu^S\\
C_\mu^A+D_\mu^S & A_\mu^A-B_\mu^A
\end{array}
\right),
\end{align}
where fields with an `$A$' (`$S$') superscript are anti-symmetric (symmetric) matrices.  
Under the $\mathbb{Z}_{2}$ symmetry, $A_\mu^A, D_\mu^S$ are even while $B_\mu^A, C_\mu^A$ are odd,  
and hence the orbifold projection sets $B_{\mu}^{A} = C_{\mu}^{A} = 0$;   
the `daughter' field is  
\begin{align}
A_{\mu}^{\rm proj}
=
i\left(
\begin{array}{cc}
A_{\mu}^A  & -D_\mu^S\\
D_\mu^S & A_\mu^A
\end{array}
\right).
\end{align} 
By using a unitary matrix
\begin{eqnarray}
P = \frac{1}{\sqrt{2}}\left(
\begin{array}{cc}
1_{N_{c}} & i 1_{N_{c}} \\
1_{N_{c}} & -i 1_{N_{c}}
\end{array} 
\right),
\end{eqnarray}
it can be rewritten as 
\begin{eqnarray}
P A_{\mu}^{\rm proj} P^{-1} =
  \left(
\begin{array}{cc}
-\mathcal{A}_{\mu}^{T} & 0\\
0 & \mathcal{A}_{\mu}
\end{array} 
\right),
\end{eqnarray}
where $\mathcal{A}_{\mu} \equiv D_{\mu}^{S} + i A^{A}_{\mu}$ is a $U(N_{c})$ gauge field.  
However, the difference between $U(N_{c})$ and $SU(N_{c})$ is a $1/N_{c}^{2}$ correction and can be neglected at large-$N_c$.~\footnote{
When one studies $U(N_c)$ theory, it is difficult to control the $U(1)$ part and the effect of the chemical potential can disappear \cite{Barbour:1986jf} 
by a lattice artifact. In order to avoid that one should simulate $SU(N_c)$ theory on the lattice. 
} The gauge part of the action after the orbifold projection is thus simply
\begin{eqnarray}
\mathcal{L^{\mathrm{gauge}, \mathrm{proj}}} = \frac{2}{4g_{SO}^{2}} \Tr \mathcal{F}_{\mu \nu}\mathcal{F}^{\mu \nu}, 
\end{eqnarray}
where $\mathcal{F}_{\mu\nu}$ is the $SU(N_{c})$ field strength. 
Let us identify it with the Lagrangian of the daughter theory times two, 
\begin{eqnarray}
\label{eq:recipe}
{\cal L}_{\rm SO} \rightarrow 2 {\cal L}_{\rm SU},
\end{eqnarray}
or equivalently  
let us take  $g_{SU}^{2} = g_{SO}^{2}$, where $g_{SU}$ is the gauge coupling constant of the $SU(N_{c})$ theory. 
This factor two is necessary in order for the ground state energies, which are proportional to the degrees of freedom, to match. 
Then expectation values of the gauge-invariant operators in parent ${\cal O}^{(p)}[A_\mu]$ agree with the expectation values of 
the daughter theory, which is obtained by replacing $A_\mu$ with $\cal A_\mu^{\rm proj}$, 
${\cal O}^{(d)}[{\cal A}_\mu]\equiv {\cal O}^{(p)}[A_\mu^{\rm proj}]$.

\begin{figure}[t]
\begin{center}
\scalebox{0.7}{
\includegraphics[width=6.5cm]{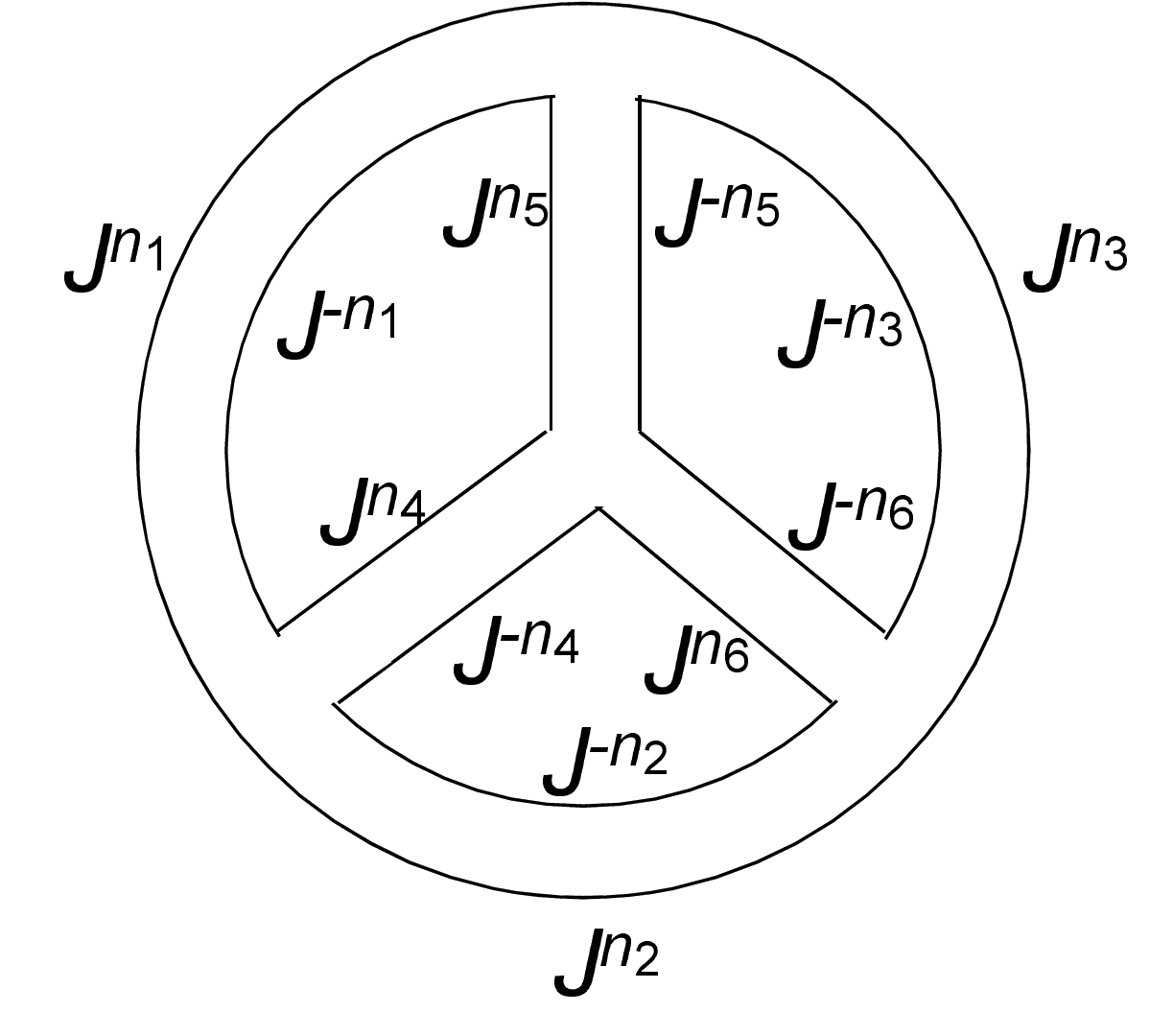}} 
\end{center}
\vspace{-0.5cm}
\caption{A vacuum planar diagram in the double-line notation.}
\label{fig:vacuum}
\end{figure}

As a pedagogical demonstration, consider a planar diagram in Fig.~\ref{fig:vacuum}.  
In order to obtain the $SU(N_c)$ diagram, we insert the projector ${\cal P}$ defined by 
\begin{eqnarray} 
A_\mu^{\rm proj}
=
{\cal P}(A_\mu^{\rm SO})
\equiv
\frac{1}{4}\sum_{n=0}^3 J_c^{n}A_\mu^{\rm SO}J_c^{-n}
=
\frac{1}{2}\left(
A_\mu^{\rm SO} + J_c A_\mu^{\rm SO}J_c^{-1}
\right). 
\end{eqnarray}
to each propagator in the $SO(2N_c)$ diagram. 
Then the only difference, if exists, comes from the contractions 
of color indices. This additional kinematic factor multiplied to the $SU(N_c)$ diagram is 
\begin{eqnarray}
\sum_{n_i=0,1}
\left(\frac{1}{2}\right)^{N_P}
\cdot \tr( J^{-n_1}J^{n_4}J^{n_5})
\cdot \tr( J^{-n_2}J^{-n_4}J^{n_6})
\cdot \tr( J^{-n_3}J^{-n_5}J^{-n_6})
\cdot \tr( J^{n_1}J^{n_2}J^{n_3}),
\end{eqnarray}
where $J=-i\sigma_2$ is a $2\times 2$ matrix and
the factor $(1/2)^{N_P}$ comes from the projectors
with $N_P=6$, where $N_P$ is the number of propagators.
Because $J$ satisfies a simple relation 
\begin{eqnarray}
Tr J^n=0
\qquad {\rm unless}
\quad 
J^n=\pm\textbf{1}_2,  
\label{eq:regularity}
\end{eqnarray}  
it is nonvanishing only when 
\begin{eqnarray}
& &
J^{-n_1}J^{n_4}J^{n_5}=\pm\textbf{1}_2,
\quad 
J^{-n_2}J^{-n_4}J^{n_6}=\pm\textbf{1}_2, 
\quad
J^{-n_3}J^{-n_5}J^{-n_6}=\pm\textbf{1}_2, 
\quad
J^{n_1}J^{n_2}J^{n_3}=\pm\textbf{1}_2,  
\nonumber\\ 
\end{eqnarray}
or equivalently 
\begin{eqnarray}
& &
-n_1+n_4+n_5={\rm even},
\quad 
-n_2-n_4+n_6={\rm even}, 
\quad
-n_3-n_5-n_6={\rm even}, 
\quad
n_1+n_2+n_3={\rm even}.   
\nonumber\\
\label{constraints}
\end{eqnarray}
Not all constraints are independent; 
actually the last one follows from the others, and hence, there are $N_L-1=3$ independent constraints, 
where $N_L=4$ is the number of index loops. 
By summing over $n_i$ for all $N_P=6$ propagators with $N_L-1=3$ constraints, 
one obtains a factor of $2^{6-3}$. 
Another factor $2^4$ comes from the  traces over color indices. 
Therefore, the total factor is 
\begin{eqnarray}
2^{-6} \cdot 2^{6-3}\cdot 2^{4}=2.
\end{eqnarray}

Generally, for given planar vacuum diagrams 
with $N_P$ propagators and $N_L$ loops,
the projectors give a factor of $(1/2)^{N_P}$,
the summation over $n_i$ under the $N_L-1$ constraints 
gives $2^{N_P-(N_L-1)}$, and the trace gives $2^{N_L}$. 
The total factor is always $2$:
\begin{eqnarray}
\label{eq:factor}
2^{-N_P} \cdot 2^{N_P-(N_L-1)} \cdot 2^{N_L}=2.
\end{eqnarray}
This factor 2 reflects the fact that the number of degrees 
of freedom in the parent theory is twice larger than that in the daughter theory.
Hence the vacuum energy per degree of freedom is equivalent between these theories.

The counting does not apply for nonplanar diagrams. Indeed, one can
easily check that the number of independent constraints is no longer $N_L-1$, 
and the factor counted in (\ref{eq:factor}) is generally different from 2 \cite{Bershadsky:1998cb}.
This is why we need to take the large-$N_c$ limit to suppress the nonplanar diagrams.

A few remarks are in order here. Firstly, in order for the orbifold equivalence to work, the projection symmetry must not be broken spontaneously. 
In this example, because the projection symmetry is embedded to the gauge transformation, it is not broken. 
In the following sections, we introduce matter fields and use flavor symmetries for the projection. 
Then the projection symmetry does break in a certain parameter region. 
Secondly, not all operators in two theories coincide. 
In the parent theory, only operators invariant under the projection symmetry is related to projected fields in the daughter. 
(In the above example this condition is not relevant because all the gauge invariant operators automatically satisfy this condition.) 
In the daughter, not all operators are obtained from the parent through the projection; in the present case, 
the projected operators are necessarily charge conjugation invariant.

\subsubsection{$Sp(2N_c)$ to $SU(N_c)$}\label{sec:Sp}
\hspace{0.51cm}
For $Sp(2N_c)$ gauge theory, 
the symplectic algebra $Sp(2N_c)$ formed by $2N_c\times 2N_c$ Hermitian matrices satisfying 
\begin{eqnarray}
J_c A^{\rm Sp} + (A^{\rm Sp})^T J_c = 0,
\end{eqnarray}
can be written using $N_c\times N_c$ matrices as
\begin{align}
A^{\rm Sp}_\mu
=
\left(
\begin{array}{cc}
iA_\mu^A + B_\mu^S & C_\mu^S - iD_\mu^S\\
C_\mu^S + iD_\mu^S & iA_\mu^A - B_\mu^S
\end{array}
\right),
\end{align}
where the fields $A_{\mu}^A$ ($B_\mu^S$, $C_\mu^S$, and $D_{\mu}^S$) 
are anti-symmetric (symmetric) matrices.  
We use the same projection condition \eqref{projection_gauge_field}; 
then one obtains $B_{\mu}^{S} = C_{\mu}^{S} = 0$ after the projection. 
This gives $SU(N_c)$ gauge theory again.

\subsection{Introducing fundamental fermions}
\hspace{0.51cm}
\subsubsection{Orbifold projection of fundamental fermions}
\hspace{0.51cm}
In this section, we introduce the orbifold projection for fundamental fermions \cite{Cherman:2010jj,Hanada:2011ju}. 

Let us consider the effect of the orbifolding on $\psi_f$. By using $\omega=e^{i\pi/2}\in U(1)_B$, we define the projection by 
\begin{eqnarray}
\psi_f = \omega J_c \psi_f. 
\label{baryon_projection}
\end{eqnarray}
 Writing 
 \begin{eqnarray}
\left(
\begin{array}{c}
\psi^{+}_{f}\\
\psi^{-}_{f}
\end{array}
\right)
= 
P\psi_{f}, 
\end{eqnarray}
the action of the $\mathbb{Z}_{2}$ symmetry is just  $(\psi^{+}_{f}, \psi^{-}_{f})\rightarrow (- \psi^{+}_{f}, \psi^{-}_{f})$. The projection consists of setting $\psi^{+}_{a} = 0$.

The action of the daughter theory is (after rescaling the coupling constant)
\begin{eqnarray}
\mathcal{L} = \frac{1}{4 g_{SU}^{2} } \Tr \mathcal{F}_{\mu \nu}^2
+ 
\sum_{f=1}^{N_f}
\bar{\psi}_{f}^{\rm SU}\left( \gamma^{\mu} {\cal D}_{\mu} + m_f + \mu\gamma^4\right)\psi_{f}^{\rm SU}, 
\end{eqnarray}
where $\mathcal{F}_{\mu\nu}$ is the field strength of the $SU(N_{c})$ gauge field $\mathcal{A}_{\mu} = D^{S}_{\mu} + i A^{A}_{\mu}$, 
$\psi_{a}^{\rm SU} = \psi^{-}_{a}$, and ${\cal D}_{\mu} = \partial_{\mu} + i \mathcal{A}_{\mu}$.  
This is an $SU(N_{c})$ gauge theory with $N_{f}$ flavors of fundamental Dirac fermions 
and the baryon chemical potential $\mu_B=\mu N_c$.   So the orbifold projection relates $SO(2N_{c})$ gauge theory to large $N_{c}$ QCD.  

On the other hand, in order to obtain fermions at finite $\mu_I$ for even $N_f$, 
we use $J_c \in SO(2N_c)$ [or $J_c \in Sp(2N_c)$] and 
$J_i \in SU(2)_{\rm isospin} [\subset SU(N_f)]$ defined by
\begin{eqnarray}
\label{eq:J_i}
J_i = - i\sigma_2 \otimes 1_{N_f/2}.
\end{eqnarray}
Here we divided 
the flavor $N_f$-component fundamental fermion is decomposed into 
two $(N_f/2)$-component fields, 
\begin{eqnarray}
\psi^{\rm SO(Sp)}
=
(\psi_i \ \psi_j), 
\end{eqnarray}
with $i$ and $j$ being the isospin indices, and $\sigma_2$ mixes $\psi_i$ and $\psi_j$.  
By using them, 
we choose the projection condition as 
\begin{eqnarray}
\label{eq:projection_isospin}
(J_c \psi^{\rm SO(Sp)} J_i^{-1})_{af}=\psi_{af}^{\rm SO(Sp)}. 
\end{eqnarray}

If we define $\varphi_{\pm}=(\psi_{\pm}^i \mp i \psi_{\pm}^j)/\sqrt{2}$
and $\xi_{\pm}=(\psi_{\pm}^i \pm \ i \psi_{\pm}^j)/\sqrt{2}$,
the fermions $\varphi_{\pm}$ survive but $\xi_{\pm}$ disappear after the projection
(\ref{eq:projection_isospin}).
Because $\varphi_{\pm}$ couple to $(A_{\mu}^{\rm SU})^C$ and $A_{\mu}^{\rm SU}$ respectively, 
the action of the daughter theory is expressed as
\begin{eqnarray}
{\cal L}_{\rm SU} = \frac{1}{4 g_{\rm SU}^{2} } \tr ({F}^{\rm SU}_{\mu \nu})^2
+ 
\sum_{f=1}^{N_f/2} \sum_{\pm}
\bar{\psi}^{\rm SU}_{f \pm}\left( \gamma^{\mu} {D}_{\mu} + m \pm \mu \gamma^4 \right)\psi^{\rm SU}_{f \pm}, 
\end{eqnarray}
where $\psi^{\rm SU}_{+}=\sqrt{2}\varphi_-$ and $\psi^{\rm SU}_{-}=\sqrt{2}\varphi_+^C$.
This theory has the isospin chemical potential $\mu_I=2\mu$.

\subsubsection{'t Hooft limit vs Veneziano limit}\label{'tHooftVsVeneziano}
\hspace{0.51cm}
In the proof of the orbifold equivalence of the pure Yang-Mills theories shown in \S~\ref{proof:pureYM}, 
the conditions \eqref{constraints} are crucial. 
What happens when the fermions are introduced? 
Firstly note that two projections \eqref{baryon_projection} and 
\eqref{eq:projection_isospin} are equivalent when the chemical potential is absent. 
Both utilize a $\mathbb{Z}_4$ subgroup of the flavor symmetry which mixes two Majorana flavors. 

Once the chemical potential is turned on, they are not equivalent anymore. 
The flavor symmetry $J_i$ used in \eqref{eq:projection_isospin} satisfies the condition similar to \eqref{eq:regularity}. 
Therefore, the proof can be repeated straightforwardly; the only difference is some color-index loops are replaced with flavor-index loops.  
On the other hand, $\mathbb{Z}_4\in U(1)_B$ used in  \eqref{baryon_projection} does not satisfy such a condition. 
Note however that one of the conditions in \eqref{constraints} is not independent and follows from other ones. 
Therefore, as long as the number of flavor-index loop is one, the proof holds. 
Because the fermion loops are suppressed by the factor $N_f/N_c$, the equivalence through \eqref{baryon_projection} holds 
in the 't Hooft large-$N_c$ limit ($N_f$ fixed) while the one through \eqref{eq:projection_isospin} holds also in the Veneziano limit ($N_f/N_c$ fixed). 

The above argument has an implication for the $1/N_c$ correction. Let us consider QCD with $\mu_B$ and that with $\mu_I$. 
In the 't Hooft large-$N_c$ limit, gluonic operators trivially agree because the fermions are not dynamical. 
Let us consider finite-$N_c$, say $N_c=3$ and $N_f=2$. Then the largest correction to the 't Hooft limit comes from 
one-fermion-loop planar diagrams, which, as we have seen,  do not distinguish $\mu_B$ and $\mu_I$. 
Therefore gluonic operators should behave similarly even quantitatively; the difference is at most $(N_f/N_c)^2$. 
In particular, the deconfinement temperatures, which is determined from 
the Polyakov loop, must be close. 
\subsubsection{Symmetry realization and validity of the equivalence}\label{sec:phase}
\hspace{0.51cm}
As we have seen so far, QCD$_B$, QCD$_I$, SO$_B$ and Sp$_B$ are   
equivalent in the large-$N_c$ limit as long as the projection symmetries are not broken spontaneously. 
In this section we discuss the phase structures of these theories and clarify when the symmetries are broken.  
It turns out that QCD$_I$, SO$_B$ and Sp$_B$ are equivalent 
everywhere in $T-\mu$ parameter space. 
The equivalence to  QCD$_B$ is not valid in the BEC/BCS crossover region of other three theories. 

Let us start with SO$_B$.  
A crucial difference from QCD is that there is no distinction between `matter' and `antimatter' because the gauge group is real.  
In other words, `fundamental' and `antifundamental' representations are equivalent. 
For this reason, mesons in this theory are not necessarily neutral under $U(1)_B$; one can construct `baryonic mesons' and 
`antibaryonic mesons' out of two `quarks' and `antiquarks', respectively.  
Because they couple to $\mu_B$, 
as the value of $\mu_B$ is increased the lightest `baryonic meson' condenses at some point.  
Then the $U(1)_B$ symmetry is broken to ${\mathbb Z}_2$ and the equivalence to QCD$_B$ fails. 
(Note that we have used ${\mathbb Z}_4$ subgroup of $U(1)_B$ for the projection.)

\begin{figure}[t]
\begin{center}
\scalebox{0.7}{
\includegraphics[width=10cm]{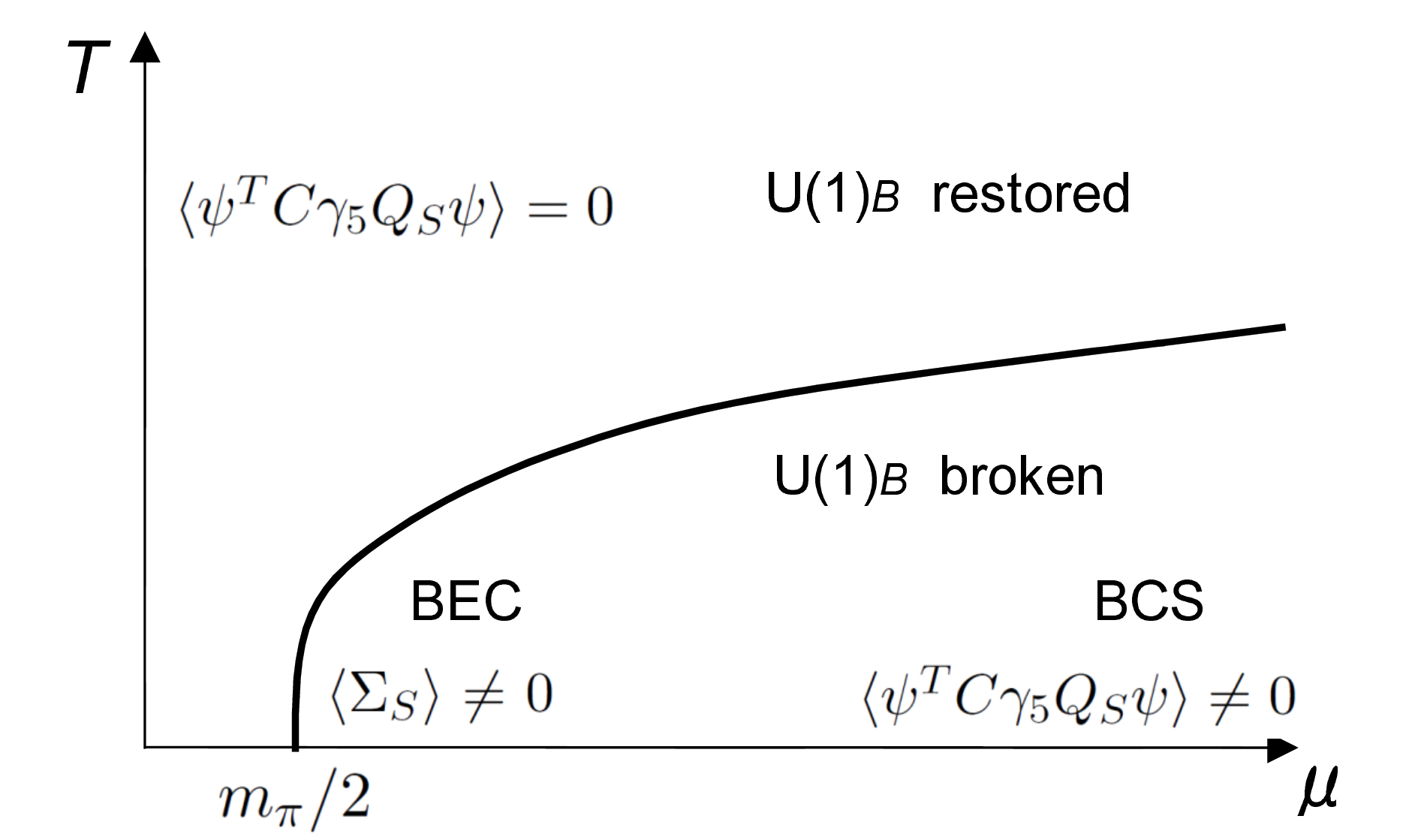}} 
\end{center}
\vspace{-0.5cm}
\caption{Phase diagram of $SO(2N_c)$ gauge theory at finite $\mu_B$. 
(Figure taken from \cite{Hanada:2011ju}.)}
\label{fig:SO}
\end{figure}

\begin{figure}[t]
\begin{center}
\scalebox{0.7}{
\includegraphics[width=10cm]{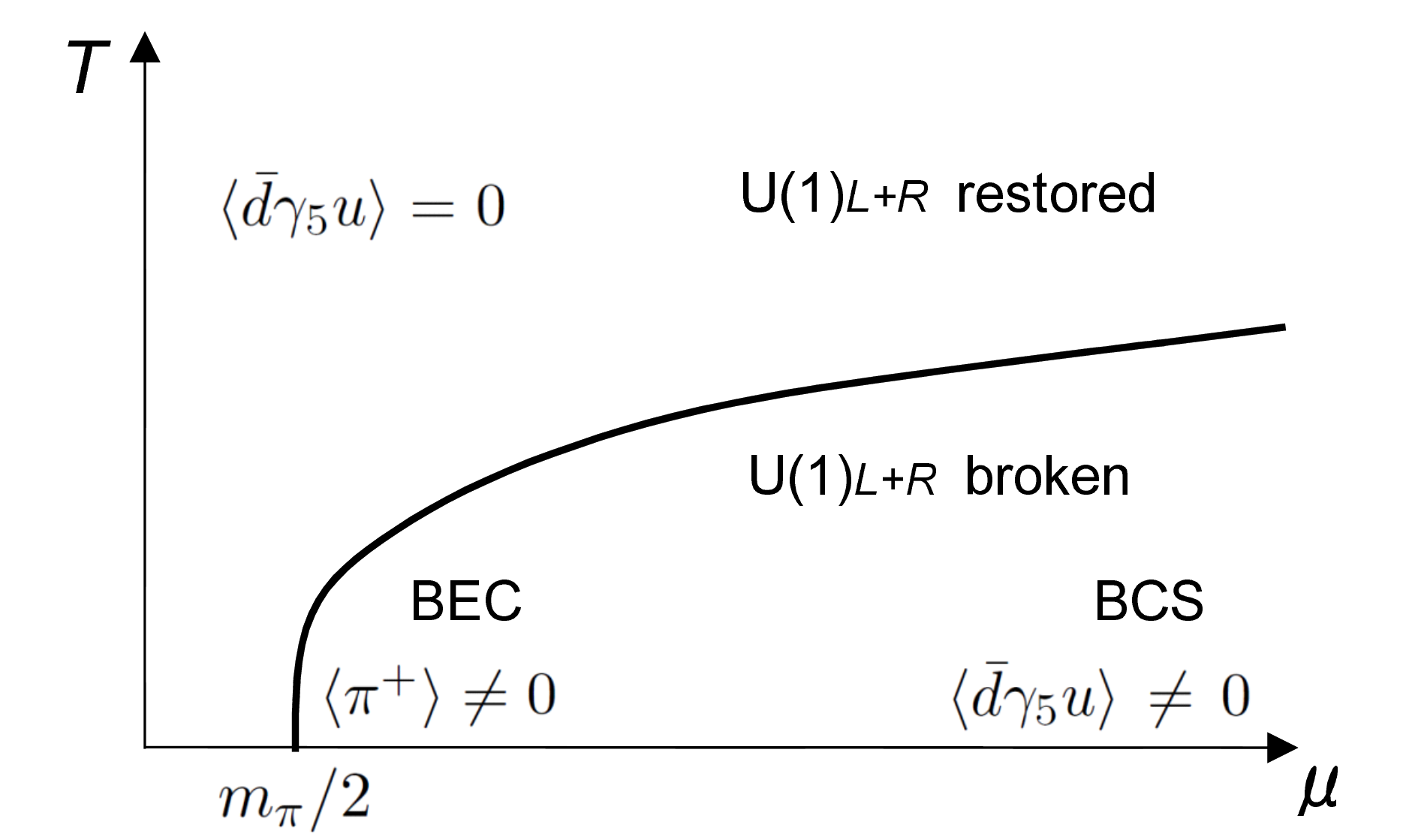}} 
\end{center}
\vspace{-0.5cm}
\caption{Phase diagram of QCD at finite $\mu_I=2\mu$. 
(Figure taken from \cite{Hanada:2011ju}.)}
\label{fig:muI}
\end{figure}

In order to identify the lightest baryonic meson, let us consider the chiral symmetry breaking in this theory. 
When $m = \mu_B = 0$, the Lagrangian \eqref{QCDlike_action} seems to have the
$SU(N_{f})_{L}\times SU(N_{f})_{R} \times U(1)_{B} \times U(1)_{A}$ 
symmetry at the classical level at first sight.  
However, chiral symmetry of the theory is known to be enhanced to $SU(2N_{f})$. 
Here $U(1)_A$ is explicitly broken by the axial anomaly.
One can actually rewrite the fermionic part of the Lagrangian (\ref{QCDlike_action})
manifestly invariant under $SU(2N_f)$
using the new variable $\Psi=(\psi_L, \sigma_2 \psi_R^*)$, which can be regarded as $2N_f$ Weyl flavors:
\begin{eqnarray}
{\cal L}_{\rm f}=i \Psi^{\dag} \sigma_{\mu} D_{\mu} \Psi,
\end{eqnarray}
where $\sigma_{\mu}=(\sigma_k,-i)$ with the Pauli matrices $\sigma_k$ ($k=1,2,3$).
The $SU(2N_f)$ chiral symmetry is spontaneously broken down to 
$SO(2N_{f})$ by the formation of the chiral condensate 
$\langle \bar{\psi}{\psi} \rangle$, leading to the $2N_f^2 + N_f -1$ Nambu-Goldstone bosons 
living on the coset space $SU(2N_{f})/SO(2N_{f})$: 
neutral pions $\Pi_a=\bar{\psi} \gamma_{5} P_a \psi$, 
`baryonic pions' (or simply `diquark')
$\Sigma_S = \psi^{T} C \gamma_5 Q_S \psi$ and 
`antibaryonic pions' $\Sigma_S^{\dag}= \psi^{\dag} C \gamma_5 Q_S \psi^*$.
It is easy to see the fate of these bosons under the orbifold projection. 
The projection to QCD$_B$ maps  neutral pions to pions in QCD, and baryonic and antibaryonic pions are projected away. 
On the other hand, the projection to QCD$_I$ sends neutral/baryonic/antibaryonic pions to 
$\pi^0$, $\pi^+$ and $\pi^-$, respectively. 
Therefore the (baryonic) pions in $SO(2N_c)$ YM and those in QCD have the same mass $m_{\pi}$. 
In the same way as the $\pi^+$ condensation in QCD$_I$ at $\mu=m_\pi/2$, 
baryonic pions condense at $\mu=m_\pi/2$ (Fig.~\ref{fig:SO} and Fig.~\ref{fig:muI}).

\begin{figure}[t]
\begin{center}
\scalebox{0.7}{
\includegraphics[width=10cm]{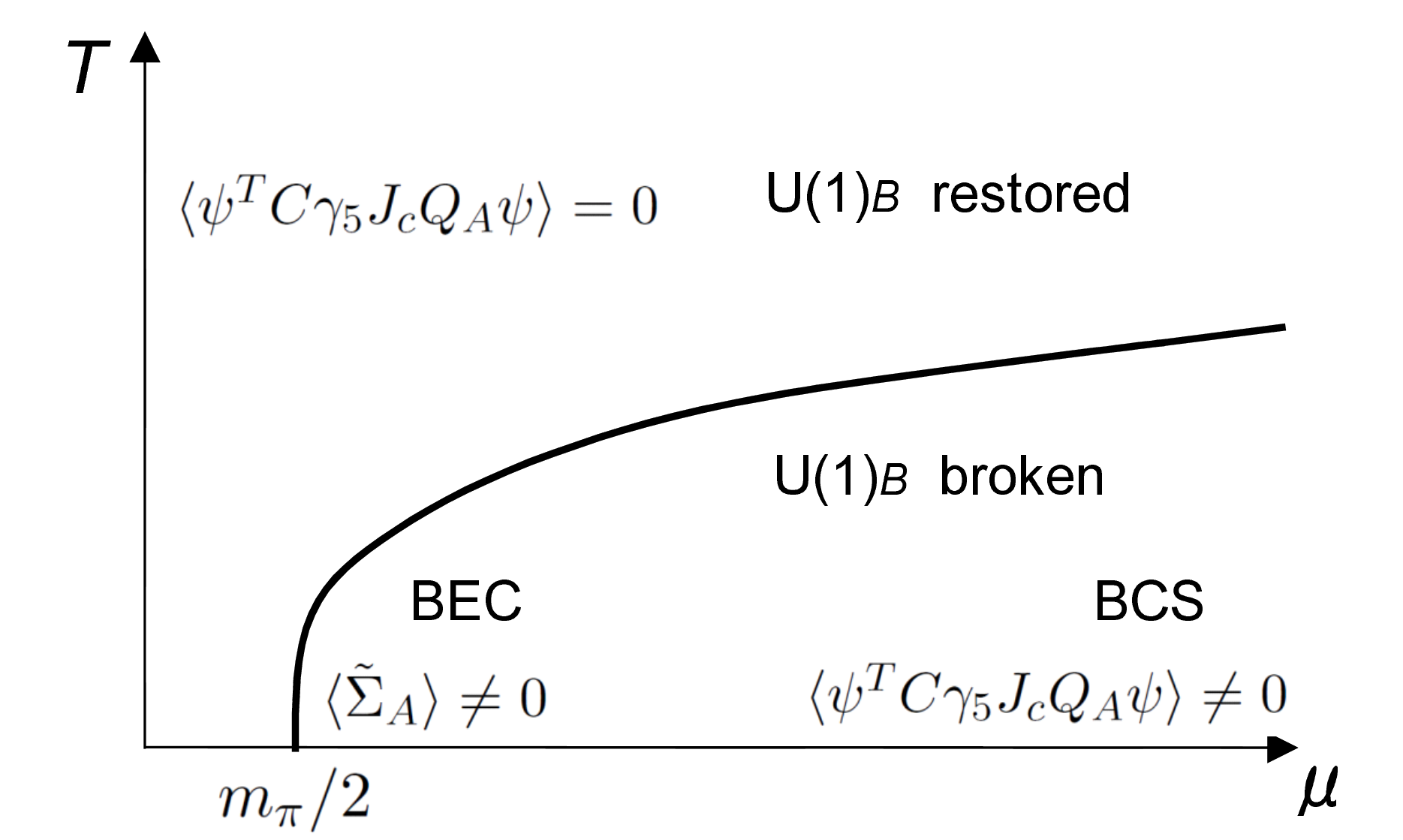}} 
\end{center}
\vspace{-0.5cm}
\caption{Phase diagram of $Sp(2N_c)$ gauge theory at finite $\mu_B$. 
$\tilde \Sigma_A = \psi^{T} C \gamma_5 J_c Q_A \psi$, where 
$Q_A$ ($A=1,2,\cdots,N_f(N_f - 1)/2$) are antisymmetric $N_f \times N_f$ matrices  in the flavor space. 
(Figure taken from \cite{Hanada:2011ju}.)
}
\label{fig:Sp}
\end{figure}

At sufficiently large $\mu$, 
the one-gluon exchange interaction in the $\psi \psi$-channel
is attractive in the color symmetric channel,
leading to the condensation of the diquark pairing $\langle \psi^{T} C \gamma_5 Q_S \psi \rangle$.
This diquark condensate does not break $SO(2N_c)$ symmetry.
This BCS pairing has the same quantum numbers 
and breaks the same $U(1)_B$ symmetry as the BEC $\langle \Sigma_S \rangle $ 
at small $\mu_B$, and there should be no phase transition for 
$\mu>m_{\pi}/2$ along $\mu$ axis.
The phase diagram of this theory is similar to that of 
QCD$_I$, as shown in Fig.~\ref{fig:SO} and Fig.~\ref{fig:muI}. 
This is because the condensates in two theories are related each other 
through the orbifold projection, and furthermore, the condensation does not break the flavor symmetry 
used for the projection.   
In the same manner, Sp$_B$and QCD$_I$ are equivalent everywhere in $T-\mu$ plane; 
see Fig.~\ref{fig:Sp}. (For further details, see \cite{Hanada:2011ju}.) 

QCD$_B$ behaves rather differently, because $\mu_B$ does not couple to mesons. 
This does not lead to a contradiction, however: because baryons are much heavier than pions, 
phenomena characteristic to QCD$_B$ (e.g. formation of hadronic matter) take place 
after the equivalence is gone due to the $U(1)_B$ breakdown in SO$_B$ and Sp$_B$ Yang-Mills 
and the pion condensation in QCD$_I$.

\section{Numerical justification at $N_c=3$}\label{sec:SU(3)}
\hspace{0.51cm}
As briefly mentioned in the introduction, the fact that the phase quenching (at $N_c=3$) is an extremely good approximation for certain quantities 
has been known empirically for long, as clearly summarized in \cite{Kogut:2007mz}. 
In this section we review previous simulation results\footnote{
In \cite{Kogut:2007mz}, as evidence for the validity of the phase quenching, they mentioned the fact that 
the expectation values of the chiral condensate take the same value in the mean-field calculation of several models. 
This is also a straightforward consequence of the orbifold equivalence; see \cite{HMY12}. 
}.

\subsection{Direct comparison by the phase reweighting}
The most straight forward way to study QCD$_B$ is the phase reweighing method, which utilizes a trivial identity 
\begin{eqnarray}
\langle{\cal O}\rangle_B
=
\frac{
\langle{\cal O}\cdot phase\rangle_I
}{
\langle phase\rangle_I,  
}
\end{eqnarray}
where $\langle\cdot\rangle_B$ and $\langle\cdot\rangle_I$ represent expectation values in QCD$_B$ and QCD$_I$, respectively, and 
$phase$ is the complex phase of the fermion determinant. 
The right hand side is calculable at least in principle. (In practice, at large volume the phase fluctuation is so violent that the average is essentialy zero.)
Because of the orbifold equivalence, for a class of observables $\langle{\cal O}\rangle_B$ and $\langle{\cal O}\rangle_I$ are the same up to $1/N_c$ correction.  
One can directly see whether this relation holds with good accuracy at $N_c=3$, at small $\mu$ and/or small volume where the phase fluctuation can be taken into 
account with a reasonable computational resource.   
(Note that a reweighting from other ensemble, say $\mu=0$, is also possible. 
However one has to choose a good ensemble which has enough overlap with QCD$_B$, 
because otherwise the importance sampling does not work. This problem is called the ``overlapping problem." 
The orbifold equivalence guarantees that the overlapping problem can be avoided by using the phase quenched ensemble (QCD$_I$). 
This provides us with the first solution to the overlapping problem with a full theoretical justification.\footnote{
The author would like to thank S.~Aoki for very fruitful discussion 
on this point.})  

In \cite{deForcrand:2007uz}, QCD$_B$ and QCD$_I$ are studied as a function of the number of up quarks, $Q$. 
The result of the former is obtained by the reweighing from QCD at $\mu=0$.
They use two staggered fermions
(corresponding to degenerate four up and four down quark species)
with the bare quark mass $m/T=0.56$ on a $8^3 \times 4$ lattice. 
In the right panel of Fig.~1 and the left panel of Fig.~4 of \cite{deForcrand:2007uz}, 
the free energy is plotted for various temperatures as functions of $Q$. 
By putting these plots on top of each other, 
one can see a very nice agreement near the critical temperature 
and $Q \lesssim 100$. It clearly shows the validity of the phase quenching. 
It should also be remarked that the corrections are still tiny for $N_f=8$;  
in the real world $N_f$, and hence the corrections, are smaller.   

In \cite{Sasai:2003py}, three-color and two-flavor QCD$_B$ and QCD$_I$ 
are studied using staggered fermions (the former is estimated by a phase reweighting from the latter). 
The bare quark mass $m/T=0.2$ and the lattice size is $8^3 \times 4$.  
The chiral condensate and the Polyakov loop are computed for $\mu/T=0.4$ and $\mu/T=0.8$, 
and QCD$_B$ and QCD$_I$ give the same value within numerical error. 

\subsection{Taylor expansion method} 
Another common approach to circumvent the sign problem is the Taylor expansion method;  
one expands the expectation value of an observable in power series of $\mu/T$, 
\begin{eqnarray}
\langle{\cal O}\rangle_B=\sum_{n=0}^\infty c_n^B\left(\frac{\mu}{T}\right)^n
\end{eqnarray}
and 
\begin{eqnarray}
\langle{\cal O}\rangle_I=\sum_{n=0}^\infty c_n^I\left(\frac{\mu}{T}\right)^n
\end{eqnarray}
in QCD$_B$ and QCD$_I$, respectively. 
Taylor coefficients $c_n^B$ and $c_n^I$, which are functions of the temperature $T$,  
can be determined by the simulation at $\mu=0$. Because of the large-$N_c$ equivalence, 
the coefficients must be the same in the large-$N_c$ limit. 

In \cite{Allton:2005gk}, the coefficients $c_2^B$ and $c_2^I$ for the chiral condensate 
and the pressure of the quark-gluon plasma have been calculated\footnote{ For odd $n$, $c_{n}^B$ 
and $c_{n}^I$ vanish, and the first nontrivial $\mu$-dependences appear in 
$c_2^B$ and $c_2^I$. Although $c_{n}^B$ ($n \geq 4$) have been calculated, 
$c_{n}^I$ ($n\ge 4$) have not been calculated in \cite{Allton:2005gk}. 
(Note that, for $n\ge 4$, they use the same symbol $c_n^I$ for another quantity.)}
in three-color and two-flavor QCD.
Their calculations are performed using staggered fermions on a $16^3 \times 4$ lattice, 
with the bare quark mass $m/T=0.4$. 
The coefficients for the pressure are \cite{Allton:2005gk} : 
\begin{center}
  \begin{tabular}{|c|c|c|}
  \hline 
$T/T_c$ &  $c_2^B$  &  $c_2^I$ \\ \hline 
0.81 & 0.0450(20) & 0.0874(8) \\
0.90 & 0.1015(24) & 0.1551(14) \\
1.00 & 0.3501(32) & 0.3822(26) \\
1.07 & 0.5824(23) & 0.5972(21) \\
1.16 & 0.7091(15) & 0.7156(14) \\
1.36 & 0.7880(11) & 0.7906(9) \\
1.65 & 0.8157(8) & 0.8169(7) \\
1.98 & 0.8230(7) & 0.8250(6) \\
  \hline
  \end{tabular}
\end{center}
Although the difference between $c_2^B$ and $c_2^I$ are not very small
for $T<T_c$ in the chiral symmetry broken (and confinement) phase, 
they agree exceptionally well for $T>T_c$. 
That the $1/N_c$ correction becomes large in the chiral symmetry broken phase is easy to understand; 
there are light modes (pions) which can easily be excited thermally. 
However they do not give a large correction to the chiral condensate; 
indeed the coefficients for the chiral condensate, which are plotted in the second panel of Fig.~3.6 of \cite{Allton:2005gk},  
shows even better agreement; actually the coefficients precisely agree within error at $T/T_c\ge 0.87$.

\subsection{Imaginary chemical potential method}
The sign problem is absent when the chemical potential 
is pure imaginary, $\mu=i\mu_{\rm img}$ ($\mu_{\rm img}\in{\mathbb R}$)
\cite{Alford:1998sd, de Forcrand:2002ci}. 
Although the imaginary chemical potential does not have a direct physical interpretation, 
it is useful if observables are analytic in $\mu^2$ around $\mu^2=0$, 
because the values at $\mu^2>0$ (real chemical potential), 
which are difficult to study due to the sign problem, may be obtained through an analytic continuation. 
Note however that the analyticity, which is necessary for the analytic continuation, 
can be lost at any phase transition, such as the chiral transition and deconfinement transition.

Our derivation for the large-$N_c$ equivalence can also 
be applied for the imaginary baryon and isospin chemical potentials, 
$(\mu_u,\mu_d)=(i\mu_{\rm img},i\mu_{\rm img})$ and 
$(\mu_u,\mu_d)=(i\mu_{\rm img},-i\mu_{\rm img})$, without any modification. 
As a result, the equivalence holds as long as the projection symmetry is unbroken. 
In \cite{Cea:2012ev}, pseudo-critical temperatures of the chiral transition  
$T_c(\mu)$ in two-flavor three-color QCD has been studied by using the imaginary chemical potential. 
(They have used the staggered fermion with the bare mass $m/T=0.2$ on a $16^3 \times 4$ lattice). 
With a quadratic ansatz
\begin{eqnarray}
\frac{T_c(\mu)}{T_c(0)}=1 + a_1 \left( \frac{\mu}{\pi T} \right)^2,
\end{eqnarray}
they found 
\begin{eqnarray}
a_1 &=& -0.465(9) \qquad {\rm for } \ \mu_I,
\nonumber \\
a_1 &=& -0.515(11) \qquad {\rm for }\ \mu_B, 
\end{eqnarray} 
which provide a nice quantitative agreement (difference$\sim$10\%) already at $N_c=3$.

\section{Demonstration of the nonperturbative equivalence in solvable models}\label{sec:analytic}
\hspace{0.51cm}
If the orbifold equivalence holds between the gauge theories,
it is natural to expect that the equivalence should hold also in solvable toy models 
which are believed to capture essential features of the gauge theories. 
In this section, we consider the chiral random matrix theory (RMT) \cite{Shuryak:1992pi}, 
and the holographic D3/D7 model  which can be studied via AdS/CFT correspondence. 
We show that the perturbative proof applies also in these cases, and furthermore, we confirm the equivalence at nonperturbative level. 
These results strongly suggest that the orbifold equivalence of the gauge theories hold nonperturbatively. 
\subsection{Chiral random matrix theories}\label{sec:RMT}
The partition function of the RMT is given by an integral over a Gaussian 
random matrix ensemble, instead of the average over the gauge field of the 
original Yang-Mills action:
\begin{eqnarray}
\label{eq:RMT}
Z=\int d\Phi \prod_{i=1}^{N_f} \det {\cal D} \ e^{-\frac{N \beta}{2}G^2 \tr \Phi^{\dagger} \Phi},
\end{eqnarray}
where $\Phi$ is an $N \times (N+\nu)$ random matrix element, 
$N$ is the size of the system, $\nu$ is the topological charge,  
and the Dirac operator ${\cal D}$ with quark mass $m_f$ is given by 
\begin{eqnarray} 
{\cal D}
=
\left(
\begin{array}{cc}
m_f \textbf{1}& \Phi+\mu\textbf{1} \\
-\Phi^\dagger +\mu\textbf{1} & m_f \textbf{1}
\end{array}
\right). 
\end{eqnarray}
We also introduced a suitable normalization with the parameter $G$ in the Gaussian.
Note that there is no spacetime coordinate in the theory; 
the size of the matrix $N$ corresponds to the spacetime volume.
It is taken to infinity in the end, corresponding to the thermodynamic limit.

Depending on the anti-unitary symmetries of the Dirac operator,
$\Phi$ is chosen as the real, complex, or quaternion real 
[see (\ref{eq:quaternion}) for the definition] matrices
denoted by the Dyson index $\beta=1$, $\beta=2$, and $\beta=4$, respectively.
The value of $\beta$ corresponds to the degrees of freedom per matrix element.
QCD and QCD-like theories corresponding to each universality class are \cite{Halasz:1997fc}: 

\begin{itemize}

\item{
$\beta=1$ : two-color QCD and $Sp(2N_c)$ gauge theory. 
 }
 
\item{
$\beta=2$ : $SU(N_c)$ QCD ($N_c\ge 3$).  
}

\item{
$\beta=4$ : $SU(N_c)$ QCD with adjoint fermions and 
$SO(N_c)$ gauge theory.  }
\end{itemize}
The effect of temperature $T$ can be incorporated as the (first) 
Matsubara frequencies by changing $\mu \rightarrow \mu + iT$ 
for one half of the determinant and $\mu \rightarrow \mu - iT$ 
for the other half of the determinant in the simplest model 
\cite{Halasz:1998qr}.  

The RMT is exactly equivalent to QCD in the $\epsilon$-regime \cite{Leutwyler:1992yt}
\begin{eqnarray}
\label{eq:low_epsilon}
\frac{1}{\Lambda_{\chi}} \ll L \ll \frac{1}{m_{\pi}}, \qquad \mu L \ll 1,
\end{eqnarray}
where $L$ is the typical scale of the system,  
because QCD reduces to a theory of zero momentum modes of pions.
In this regime, the system has a universality, i.e., the dynamics depends 
only on the symmetry breaking pattern and is independent of the microscopic details;
QCD can be replaced by the RMT with the same global symmetry breaking pattern.
Outside the $\epsilon$-regime, the universality is lost. However, the RMT is 
still useful as a schematic model to study the qualitative properties of QCD 
such as the phase structure at finite $T$ and $\mu$ \cite{Halasz:1998qr}.
The advantage of the RMT is that it can be solved analytically 
although QCD itself cannot be. 

As mentioned above, the RMT has the size of the matrix $N$, 
which should be taken to infinite (thermodynamic limit) in the end.
In this sense, the RMT is a ``large-$N$" matrix model, and hence, 
the perturbative proof of the orbifold equivalence given in Sec.~\ref{sec:equivalence} is 
directly applicable. 
Note that the size of the random matrix is not related to the number of color $N_c$.
The RMT is analytically solvable and hence the nonperturbative orbifold equivalence can be checked explicitly.  
\subsubsection{Orbifold projections in the chiral random matrix theories} 
\label{sec:RMT_equivalence}
In this section, we construct the orbifold projections in the 
chiral random matrix theories (RMTs) between $\beta=4$, $\beta=2$, and $\beta=1$.
Thereby a class of observables in the RMTs with the different Dyson indices
are found to be identical to each other.
In the following, we will concentrate on the RMT at finite $\mu$ and $T=0$, which can be easily generalized to nonzero $T$.
For simplicity, we set $\nu=0$ and quark masses to be common, $m_f=m$.

\begin{figure}[t]
\begin{center}
\scalebox{0.7}{
\includegraphics[width=11cm]{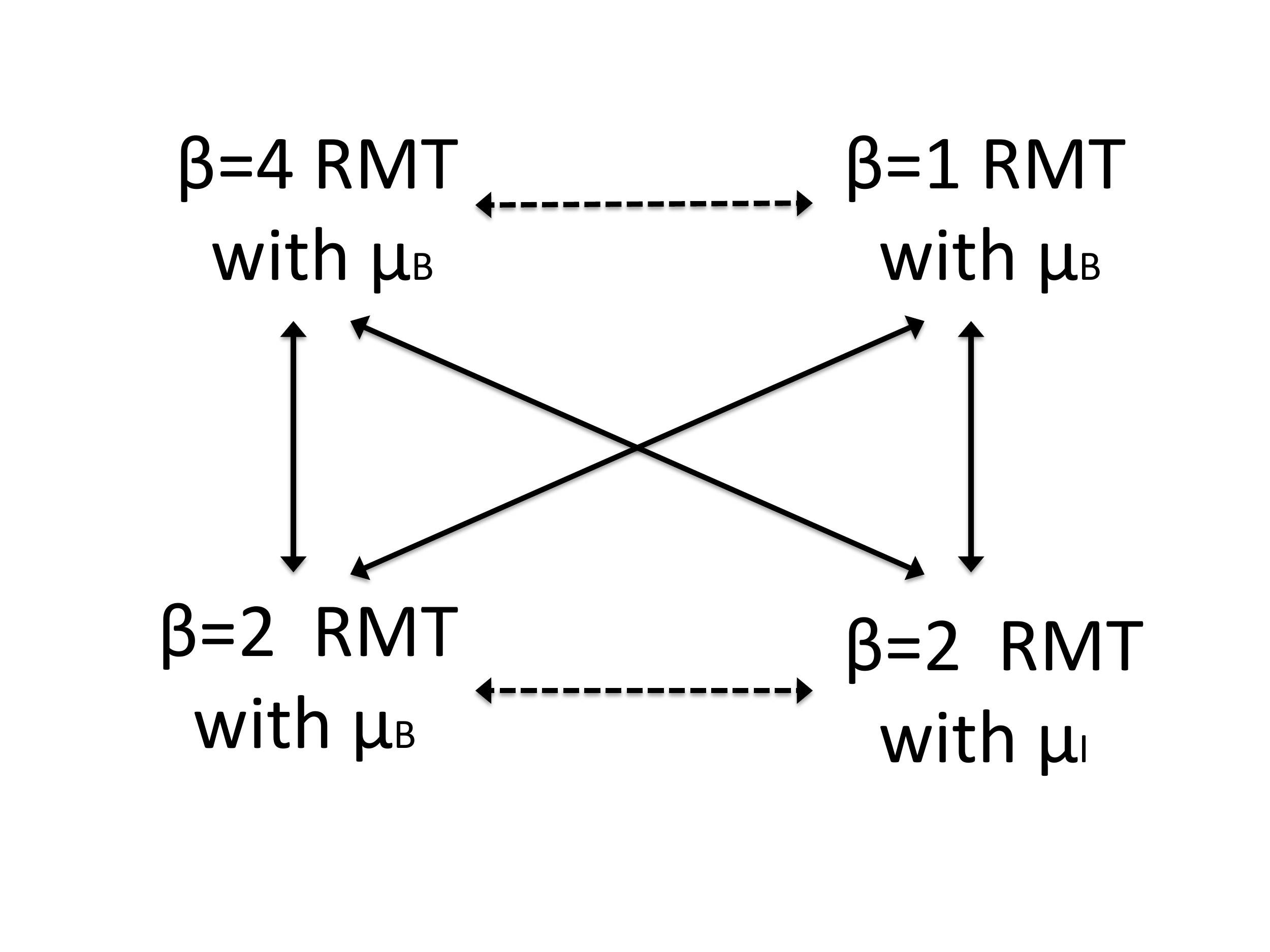}}
\end{center}
\vspace{-0.5cm} 
\caption{Relations between $\beta=2$ RMT at finite $\mu_B$ or $\mu_I$
and $\beta=4$ and $\beta=1$ RMTs at finite $\mu_B$ through orbifolding. 
$\beta=2$ RMT at small and large $\mu_I$ 
can be obtained from $\beta=4$ and $\beta=1$ RMTs at small and large $\mu_B$.
$\beta=2$ RMT at small $\mu_B$ can also be obtained from
$\beta=4$ RMT at small $\mu_B$, while $\beta=2$ RMT at large $\mu_B$
inside the BEC-BCS crossover region cannot.}
\label{fig:RMT}
\end{figure}

The relationships between these RMTs via orbifold projections are summarized 
in Fig.~\ref{fig:RMT}.  
We start with the $\beta=4$ or $\beta=1$ RMT at finite $\mu_B$
with the size of $\Phi$ being $2N$, 
and define the orbifold projection to the $\beta=2$ RMT
at finite $\mu_B$ or $\mu_I$ with the size $N$.

The action of the $\beta=4$ RMT is given by 
\begin{eqnarray}
Z=\int d\Phi d\Psi \ e^{-S}, \qquad S=S_{B}+S_{F},
\end{eqnarray}
where 
\begin{eqnarray}
\label{eq:RMT-action}
S_B
=\frac{N\beta}{2}G^2 \tr \Phi^\dagger \Phi, \qquad
S_F
=
\sum_{f=1}^{N_f}
\bar{\Psi}_f {\cal D}\Psi_f, 
\end{eqnarray}
and  
\begin{eqnarray}
{\cal D}
=
\left(
\begin{array}{cc}
m \textbf{1}_{2N}& \Phi+\mu\textbf{1}_{2N} \\
-\Phi^\dagger +\mu\textbf{1}_{2N} & m \textbf{1}_{2N}
\end{array}
\right).   
\end{eqnarray}
Here $\Psi_f$ are complex Grassmann $4N$-component vectors and
$\Phi$ is a $2N\times 2N$ quaternion real matrix of the form:
\begin{eqnarray}
\label{eq:quaternion}
\Phi
\equiv \sum_{\mu=0}^3 a^{\mu} i \sigma_{\mu}=
\left(
\begin{array}{cc}
a^0 + i a^3 &  a^2 + i a^1 \\
-a^2 + i a^1 & a^0 - i a^3
\end{array}
\right), 
\end{eqnarray}
where $a^{\mu}$ are $N\times N$ real matrices 
and $\sigma_{\mu}=(-i, \sigma_k)$ with Pauli matrices $\sigma_k$.

For the bosonic matrix $\Phi$, we impose the projection condition 
\begin{eqnarray}
J_c \Phi J_c^{-1} = \Phi, \qquad
J_c \equiv
\left(
\begin{array}{cc}
& -\textbf{1}_{N}   \\
\textbf{1}_{N}  & 
\end{array}
\right).   
\end{eqnarray}
Then we obtain 
\begin{eqnarray}
\Phi^{\rm proj}
=
\left(
\begin{array}{cc}
a^0  & a^2\\
-a^2 & a^0 
\end{array}
\right), 
\end{eqnarray}
which is equivalent to two copies of a $N\times N$ complex matrix
after a unitary transformation
\begin{eqnarray}
P \Phi^{\rm proj} P^{-1} =
  \left(
\begin{array}{cc}
\phi^\ast& 0\\
0 & \phi
\end{array} 
\right) 
\equiv \Phi_{\beta=2}, \qquad
P \equiv \frac{1}{\sqrt{2}}\left(
\begin{array}{cc}
\textbf{1}_{N} & i \textbf{1}_{N} \\
\textbf{1}_{N} & -i \textbf{1}_{N}
\end{array} 
\right),
\end{eqnarray}
where $\phi = a^0+ia^2$.  
The bosonic part of the action is mapped to the one for the 
$\beta=2$ RMT. 
Note that the factor 2 in  (\ref{eq:recipe}) is
reflected in the difference of normalization between $\beta=4$ and $\beta=2$
in (\ref{eq:RMT-action}) if the trace for $\beta=4$ is defined as 
the so-call ``QTr" which is one-half the usual trace.

In order to define a projection for the fermions, we write $\Psi$ by using two
$2N$-component fermions $\psi_R$ and $\psi_L$ as
\begin{eqnarray}
\Psi=
\left(
\begin{array}{c}
\psi_R \\
\psi_L
\end{array}
\right). 
\end{eqnarray}
Here $\psi_R$ and $\psi_L$ are further decomposed into two $N$-component fermions
\begin{eqnarray}
\psi_R=
\left(
\begin{array}{c}
\psi_R^1\\
\psi_R^2
\end{array}
\right), \qquad
\psi_L=
\left(
\begin{array}{c}
\psi_L^1\\
\psi_L^2
\end{array}
\right),
\end{eqnarray}
where the flavor index is suppressed for simplicity. 
Then it is straightforward to check that the projection 
\begin{eqnarray}
\psi_R = \omega J_c \psi_R,
\qquad
\psi_L = \omega J_c \psi_L,  
\end{eqnarray}
where $\omega=e^{i\pi/2}$ as before, 
provides us with the $\beta=2$ RMT at finite baryon chemical potential. 
In a similar manner, the $\beta=2$ RMT at finite isospin chemical potential 
can be obtained by imposing the projection condition
\begin{eqnarray}
J_c\psi_R J_i^{-1}=\psi_R, \qquad J_c\psi_L J_i^{-1}=\psi_L.
\end{eqnarray}

The $\beta=2$ RMT can also be obtained from the $\beta=1$ RMT.
We start with the action of the $\beta=1$ RMT given by
(\ref{eq:RMT-action}), but this time 
$\Phi$ is a $2N\times 2N$ real matrix,  
which can be parametrized as
\begin{eqnarray}
\Phi =
\left(
\begin{array}{cc}
a^0 +  a^3 &  a^2 +  a^1 \\
-a^2 +  a^1 & a^0 -  a^3
\end{array}
\right), 
\end{eqnarray}
where $a^{\mu}$ are $N\times N$ real matrices. Note that the only change 
in this expression compared with (\ref{eq:quaternion}) is that the 
factors $i$ in front of $a^0$ and $a^3$ are absent.
Then one can easily find that the same projection conditions for $\Phi$
and $\Psi$ in the previous subsection gives the $\beta=2$ RMT 
at finite $\mu_B$ or finite $\mu_I$.

\subsubsection{Explicit demonstration of the nonperturbative equivalence} 
\label{sec:solveRMT}
The orbifold equivalence in the RMT predicts that the $\beta=4$ and $\beta=1$ RMTs
at finite $\mu_B$ and $\beta=2$ RMT at finite $\mu_I$ are equivalent to each other for any $m, T$ and $\mu$ 
in the neutral sector, to all order in perturbation theory. Outside the (baryonic) pion condensation phase, 
the above three theories must also be equivalent to the $\beta=2$ RMT at finite $\mu_B$. 
Below we verify the nonperturbative equivalence by computing the effective potential of each RMT explicitly.  

For the $\beta=2$ and $\beta=1$ RMTs, the effective potentials are computed in 
\cite{Klein:2003fy, Klein:2004hv}, and the coincidence of the effective potential
of the $\beta=2$ RMT at finite $\mu_B$ (at $\mu_I=0$) and that of the $\beta=2$ RMT at finite $\mu_I$ 
(at $\mu_B=0$) outside the pion condensed phase has been pointed out.
Below we first summarize the effective potential of the $N_f=2$ RMT  
with the quark mass $m_f$ and the chemical potential 
$\mu_f$ for each flavor, $f=1,2$. 
The average baryon and isospin chemical potentials $\bar{\mu}_{B}$ and $\bar{\mu}_I$ are defined as
\begin{eqnarray}
\bar \mu_B &\equiv & \frac{\mu_B}{N_c} = \frac{1}{2}(\mu_1 + \mu_2),
\\
\bar \mu_I &\equiv & \frac{\mu_I}{2} = \frac{1}{2}(\mu_1 - \mu_2).
\end{eqnarray}
We denote the parameters of the RMT representing 
the chiral condensate $\sigma_f$, pion condensate $\rho$, and 
diquark (baryonic pion) condensate $\Delta$, and their sources as
$m_f$, $\lambda$, and $J$, respectively. 
Then we show that the equivalence holds nonperturbatively between RMTs
as a natural consequence of the orbifold projections. 
The importance of the unbroken projection symmetry will become clear through the argument. 
\\
\\
\textbf{Effective potential of $\beta=4$}
\\
We first consider the $\beta=4$ RMT with degenerate quark mass $m_f=m$ 
at finite baryon chemical potential $\mu_f = \bar \mu_B$.  
The effective potential is  \cite{Hanada:2011ju}
\begin{eqnarray}
\label{eq:potential4}
\Omega_{\beta=4} = 16G^2\left[ \left(\sigma- \frac{m}{2} \right)^2 
+ \left(\Delta- \frac{J}{2} \right)^2 \right] 
- 2 \sum_{\pm} \ln [4\sigma^2 + 4\Delta^2 - (\bar \mu_B \pm iT)^2].
\end{eqnarray}
The chiral condensate and the diquark condensate are expressed using 
$\sigma$ and $\Delta$ as
\begin{eqnarray}
\langle \bar u u \rangle_{\beta=4} 
&=& \left. \frac{1}{4N}\partial_{m}\ln Z_{\beta=4} \right|_{m=0} = - 4G^2 \sigma_{\beta=4},
\\
\langle \bar u^T C \gamma_5 u \rangle_{\beta=4} 
&=& \left. \frac{1}{4N}\partial_{J}\ln Z_{\beta=4} \right|_{J=0} = - 4G^2 \Delta_{\beta=4}.
\end{eqnarray}
\\
\\
\textbf{Effective potential of $\beta=2$}
\\
The effective potential of $\beta=2$ is \cite{Klein:2003fy} 
\begin{eqnarray}
\label{eq:potential2}
\Omega_{\beta=2} &=& G^2[(\sigma_1-m_1)^2+(\sigma_2-m_2)^2 + 2(\rho-\lambda)^2]
\nonumber \\
	& & -\frac{1}{2}\sum_{\pm}
\ln [(\sigma_1 + \mu_1 \pm iT)(\sigma_2 - \mu_2 \mp iT) + \rho^2]
[(\sigma_1 - \mu_1 \mp iT)(\sigma_2 + \mu_2 \pm iT) + \rho^2].
\nonumber \\
\end{eqnarray}
The chiral condensate and pion condensate are  
\begin{eqnarray}
\langle \bar u u \rangle_{\beta=2} 
&=& \left. \frac{1}{2N}\partial_{m_1}\ln Z_{\beta=2} \right|_{m_1=0} = - G^2 \sigma_{\beta=2},
\\
\langle \bar d \gamma^5 u \rangle_{\beta=2} 
&=& \left. \frac{1}{4N}\partial_{\lambda}\ln Z_{\beta=2} \right|_{\lambda=0} = - G^2 \rho_{\beta=2}.
\end{eqnarray}
Note that, as long as $\rho=0$ (i.e., outside the pion condensed phase), 
the potential (\ref{eq:potential2}) is a function of $\mu_1^2=(\bar \mu_B + \bar \mu_I)^2$ 
and $\mu_2^2 = (\bar \mu_B - \bar \mu_I)^2$. Therefore,  
\begin{eqnarray}
\label{eq:equivalence2}
\Omega_{\beta=2}(\bar \mu_B=\mu, \bar \mu_I=0)
=\Omega_{\beta=2}(\bar \mu_I=\mu, \bar \mu_B=0) \ \ {\rm for} \ \ {\rho=0}.
\end{eqnarray}
Here $\rho=0$ means that the projection symmetry,
which is used for the orbifolding in Sec.~\ref{sec:RMT_equivalence}, 
is not broken.
From (\ref{eq:equivalence2}), the magnitude of the chiral condensate $\sigma$ and 
the critical temperature of chiral phase transition $T^{\sigma}$ in each theory coincide,
\begin{eqnarray}
\sigma_{\beta=2}(\bar \mu_B)|_{\bar \mu_I=0}
&=&\sigma_{\beta=2}(\bar \mu_I)|_{\bar \mu_B=0} \ \ {\rm for} \ \ {\rho=0},
\\
T^{\sigma}_{\beta=2}(\bar \mu_B)|_{\bar \mu_I=0}
&=&T^{\sigma}_{\beta=2}(\bar \mu_I)|_{\bar \mu_B=0} \ \ {\rm for} \ \ {\rho=0},
\end{eqnarray}
as a consequence of the orbifold equivalence.
Especially, this shows that the phase-quenched approximation for $\sigma_{\beta=2}(\bar \mu_B)$
and $T^{\sigma}_{\beta=2}(\bar \mu_B)$ works outside the pion condensed phase, 
as mentioned in \cite{Klein:2003fy}. 

Note that, 
even though the effective potentials are identical for $\rho=0$ in (\ref{eq:equivalence2}),
the partition functions themselves are not generally the same. This is because the 
pre-exponential factor also contributes to the partition function, 
which is not taken into account in computing the effective potential.  
Therefore, the sign problem measured as the phase of the partition function
can be severe inside as well as outside the pion condensed phase \cite{Han:2008xj}.  
The result here shows that the phase-quenched approximation is exact for the observables 
above independently of the severity of the sign problem, as long as $\rho=0$. 
\\
\\
\textbf{Effective potential of $\beta=1$}
\\
The effective potential of $\beta=1$ is \cite{Klein:2004hv} 
\begin{eqnarray}
\label{eq:potential1}
\Omega_{\beta=1} &=& G^2[(\sigma_1-m_1)^2+(\sigma_2-m_2)^2 + 2(\rho-\lambda)^2 + 2(\Delta-J)^2]
\nonumber \\
	& & -\frac{1}{4}\sum_{\pm}
\ln \{[(\sigma_1 + \mu_1 \pm iT)(\sigma_2 - \mu_2 \mp iT) + \rho^2 + \Delta^2 ]
\nonumber \\
& & \qquad \qquad \times[(\sigma_1 - \mu_1 \pm iT)(\sigma_2 + \mu_2 \mp iT) + \rho^2 + \Delta^2] + 4\Delta^2 \mu_1 \mu_2 \}
\nonumber \\
& & \qquad \qquad \times \{[(\sigma_1 - \mu_1 \mp iT)(\sigma_2 + \mu_2 \pm iT) + \rho^2 + \Delta^2 ]
\nonumber \\
& & \qquad \qquad \times (\sigma_1 + \mu_1 \mp iT)(\sigma_2 - \mu_2 \pm iT) + \rho^2 + \Delta^2] + 4\Delta^2 \mu_1 \mu_2 \}.
\end{eqnarray}
The chiral condensate, pion condensate, and diquark condensate are
\begin{eqnarray}
\langle \bar u u \rangle_{\beta=1} 
&=& \left. \frac{1}{2N}\partial_{m_1}\ln Z_{\beta=1} \right|_{m_1=0} = - G^2 \sigma_{\beta=1},
\\
\langle \bar d \gamma^5 u \rangle_{\beta=1} 
&=& \left. \frac{1}{4N}\partial_{\lambda}\ln Z_{\beta=1} \right|_{\lambda=0} = - G^2 \rho_{\beta=1}.
\\
\langle d^T C \gamma_5 u \rangle_{\beta=1} 
&=& \left. \frac{1}{4N}\partial_{J}\ln Z_{\beta=1} \right|_{J=0} = - G^2 \Delta_{\beta=1}.
\end{eqnarray}
The potential (\ref{eq:potential1}) has the symmetry
\begin{eqnarray}
\Omega_{\beta=1}(\Delta, \rho, \mu_1, \mu_2)
=\Omega_{\beta=1}(\rho, -\Delta, \mu_1, -\mu_2),
\end{eqnarray}
due to the $\bar \mu_B \leftrightarrow \bar \mu_I$ symmetry for $\beta=1$.
Note that this symmetry has nothing to do with the orbifold equivalence.
\\
\\
\textbf{Nonperturbative orbifold equivalence between $\beta=4$, $\beta=2$, and $\beta=1$}
\\  
By comparing (\ref{eq:potential4}), (\ref{eq:potential2}), and (\ref{eq:potential1}),
and by using the $\bar \mu_B \leftrightarrow \bar \mu_I$ symmetry for $\beta=1$, 
one finds the relation 
(note that $\Delta=0$ at $\bar \mu_B=0$ and $\rho=0$ at $\bar \mu_I=0$):
\begin{eqnarray}
\label{eq:EP}
\Omega_{\beta=4}(2\sigma_{\beta=4},2\Delta_{\beta=4})|_{\bar \mu_B=\mu, \bar \mu_I=0}
&=& 2\Omega_{\beta=2}(\sigma_{\beta=2},\rho_{\beta=2})|_{\bar \mu_I=\mu, \bar \mu_B=0}
\nonumber \\
&=& 2\Omega_{\beta=1}(\sigma_{\beta=1},\Delta_{\beta=1})|_{ \bar \mu_B=\mu, \bar \mu_I=0}.
\end{eqnarray}
Unlike the relation (\ref{eq:equivalence2}), this
is valid not only for $\rho=0$ (or $\Delta=0$) 
but also for $\rho \neq 0$ (or $\Delta \neq 0$). 
This is expected because the condensation does not break the projection symmetry, as discussed 
in \S~\ref{sec:equivalence}. 
The factor 2 of the effective potentials, mentioned in (\ref{eq:recipe}), 
comes from the fact that the $\beta=4$ RMT with the size of $\Phi$ being $2N$
twice more degrees of freedom compared to the $\beta=2$ or $\beta=1$ RMT with the size $N$.
The origin of the factor 2 for the arguments of 
the potential of the $\beta=4$ RMT is the same; fermions in $\beta=4$ theory has twice larger degrees of freedom, 
$\langle \bar u u \rangle_{\beta=4}=2\langle \bar u u \rangle_{\beta=2}
=2\langle \bar u u \rangle_{\beta=1}$.
The relation (\ref{eq:EP}) leads to the coincidence of the magnitudes of the order parameters
(up to the factor 2) and the critical temperatures, 
\begin{eqnarray}
2\sigma_{\beta=4}(\bar \mu_B=\mu)|_{\bar \mu_I=0}
=\sigma_{\beta=2}(\bar \mu_I=\mu)|_{\bar \mu_B=0}
=\sigma_{\beta=1}(\bar \mu_B=\mu)|_{\bar \mu_I=0},
\\
2\Delta_{\beta=4}(\bar \mu_B=\mu)|_{\bar \mu_I=0}
=\rho_{\beta=2}(\bar \mu_I=\mu)|_{\bar \mu_B=0}
=\Delta_{\beta=1}(\bar \mu_B=\mu)|_{\bar \mu_I=0},
\\
T^{\sigma}_{\beta=4}(\bar \mu_B=\mu)|_{\bar \mu_I=0}
=T^{\sigma}_{\beta=2}(\bar \mu_I=\mu)|_{\bar \mu_B=0}
=T^{\sigma}_{\beta=1}(\bar \mu_B=\mu)|_{\bar \mu_I=0},
\\
T^{\Delta}_{\beta=4}(\bar \mu_B=\mu)|_{\bar \mu_I=0}
=T^{\rho}_{\beta=2}(\bar \mu_I=\mu)|_{\bar \mu_B=0}
=T^{\Delta}_{\beta=1}(\bar \mu_B=\mu)|_{\bar \mu_I=0},
\end{eqnarray}
which are expected 
as a consequence of the orbifold equivalence.
We note that, 
the equivalence of the neutral order parameters and the
critical temperatures should be satisfied in the original QCD and QCD-like theories
as we claimed in \S\ref{sec:equivalence}, while 
the effective potentials will not necessarily coincide in QCD.
The RMT has much less degrees of freedom;  the effective potential is a 
function of only the neutral order parameters,  and furthermore, 
all the moments are identical due to the orbifold equivalence. 
As a result, the effective potentials must be identical. 
In QCD and QCD-like theories, the effective potentials depend also on 
non-neutral observables, and hence the effective potentials are not identical in general.

\subsection{Holographic models}\label{sec:holography}
\begin{figure}[htbp]
\begin{center}
\scalebox{0.3}{
\rotatebox{0}{
\includegraphics{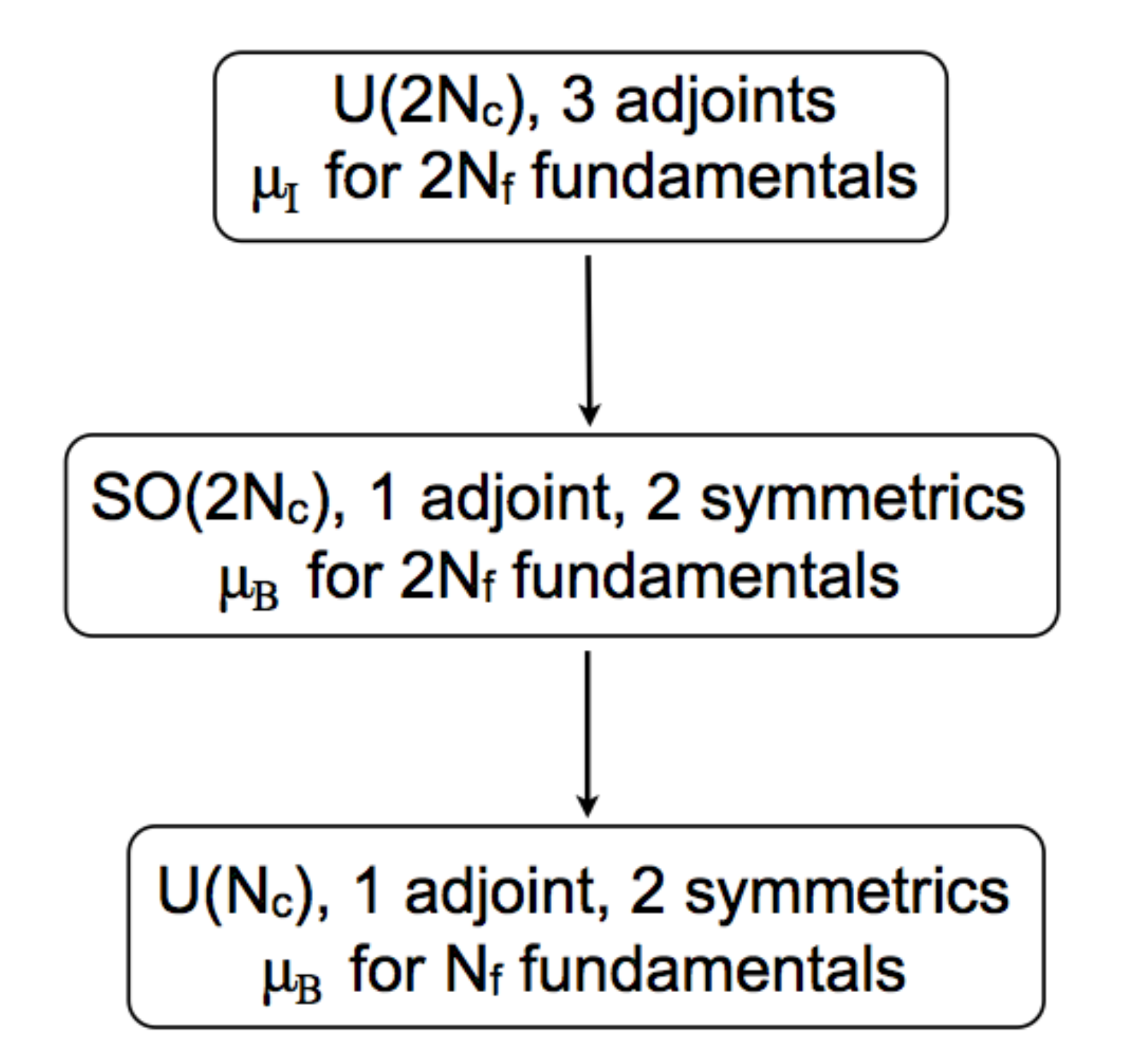}}}
\caption{
${\cal N}=1$ Supersymmetric version of the orbifold projection which has a holographic realization. 
}\label{fig:Projection_SYM}
\end{center}
\end{figure}

\hspace{0.51cm}
Among other interesting toy models are supersymmetric analogues of large-$N_c$ QCD which have gravity dual description. 
In  \cite{Hanada:2012nj} 
the D3/D7 system with chemical potential has been studied (Fig.~\ref{fig:Projection_SYM}). The starting point is 4d ${\cal N}=4$ $U(2N_c)$ super Yang-Mills, 
which is dual to type IIB supergravity in $AdS_5\times S^5$. We introduce $N_f$ D7 branes winding on three-cycle of $S^5$. 
Then open strings stretching between D3 and D7 can be regarded as `quarks' with 'flavor symmetry' $U(N_f)$. 
In the 't Hooft limit $N_f/N_c\to 0$, D7-branes behave as probes, and their dynamics is described by the Dirac-Born-Infeld (DBI) action 
on the $AdS_5\times S^5$ background. 
The isospin chemical potential $\mu_I$ can be introduced through the boundary condition of the gauge field in the DBI action. 
This theory can be projected to an $SO(2N_c)$ theory with $\mu_B$ through an orientifold projection, and further down to $U(N_c)$ theory with $\mu_B$. 
However because the orientifold projection does not change the local structure of the brane configuration, 
the equation of motion remains untouched as long as the projection symmetry is not broken spontaneously.  
Therefore, the dynamics of mesons determined by the DBI action coincide, and the orbifold equivalence follows.  
Note that this is a `nonperturbative proof', if we assume the AdS/CFT duality holds nonperturbatively. 

Actually these theories are solved in \cite{Mateos:2007vc} ($U(N_c)$ with $\mu_B$) and \cite{Erdmenger:2008yj,Ammon:2008fc} ($U(N_c)$ with $\mu_I$), 
and as long as the necessary symmetry is intact the equivalence can be seen explicitly. 

\section{Conclusion and outlook} \label{sec:conclusion}
\hspace{0.51cm}
 We have pointed out that QCD and  various QCD-like theories with chemical potentials are equivalent at large-$N_c$ 
 through the orbifold equivalence, at least to all order in perturbation theory. 
 QCD$_I$, SO$_B$ and Sp$_B$ are equivalent everywhere in the $T$-$\mu$ plane, and furthermore, they are equivalent to 
 QCD$_B$ outside the BEC-BCS crossover region. 
 
Our result has immediate implication for the study of the chiral and deconfinement transitions in high-$T$, small-$\mu$ region. 
In this region it is reasonable to assume the $N_f/N_c$ expansion is not bad   
and hence we can expect that 
the Monte-Carlo results of the QCD with isospin chemical potential describe the QCD with the baryon chemical potential 
with rather good accuracy. 
(Note that the deviation is only $O((N_f/N_c)^2)$, as explained in \S~\ref{'tHooftVsVeneziano}.) 
Indeed, as reviewed in \ref{sec:SU(3)}, all previous simulation results confirmed the validity of the phase quenching at $N_c=3$. 
At small volume, it is possible to take into account the $1/N_c$ corrections by the phase reweighting method. 
Because of the large-$N_c$ equivalence, many important observables are free from the overlapping problem 
(in other words the overlapping problem is $1/N_c$ suppressed.) 

Furthermore, by using the SO$_B$, one can study three-flavor theory without suffering from the sign problem.  
In a similar manner, from simulation results of the two-color QCD and adjoint QCD, which belong to the same universality classes 
as $Sp(2N_c)$ and $SO(2N_c)$ theories, respectively, 
one can extract qualitative information relevant for $SU(3)$ QCD$_B$. \footnote{
Strictly speaking `universality class' is meaningful only in the $\varepsilon$-regime, while the large-$N_c$ limit with fixed quark mass, 
chemical potential and volume is not in that regime. Therefore the similarity is expected only at qualitative level, similarly 
to the case of the chiral random model in the planar limit discussed in \S~\ref{sec:RMT_equivalence}. 
We thank N.~Yamamoto, J.~Verbaarschot and P.~H.~Damgaard for discussions on this issue. 
}  
Similar study e.g. phase quenched simulation of $SU(3)$ 3-flavor QCD has been performed \cite{Kogut:2007mz}  
and the result suggests the QCD critical point does not exist. 
Therefore it is very important to study these sign-free theories numerically, further in detail,  
in order to find (or exclude) the QCD critical point.

\section*{Acknowledgement} 
\hspace{0.51cm}
The author would like to thank Aleksey Cherman, Yoshinori Matsuo, Daniel Robles-Llana and Naoki Yamamoto for fruitful collaborations which this paper is based on. 
He thank Naoki Yamamoto also for valuable comments on this paper at various stages. 
He also thanks Philippe de Forcrand, Shoji Hashimoto, Carlos Hoyos, Andreas Karch, Keitaro Nagata, Atsushi Nakamura, 
Brian Tiburzi and Laurence Yaffe for stimulating discussions, comments, and/or related collaborations. 
The author was in part supported by Japan Society for the Promotion of Science Postdoctoral Fellowships for Research Abroad.

\appendix


\end{document}